\documentclass[useAMS,usenatbib]{mn2e}
\usepackage{aas_macros}
\usepackage{amsmath,amssymb}
\usepackage{bm}
\usepackage{color}
\usepackage[dvipdfmx]{hyperref}
\usepackage{graphicx}
\usepackage{ulem}

\newcommand{\mr}[1]{\mathrm{#1}}
\newcommand{\cmr}[1]{\,\mathrm{#1}}

\usepackage{amsmath}	

\begin{document}
\author[K. Sugimura et al.
]{Kazuyuki Sugimura,$^1$\thanks{E-mail: sugimura@astr.tohoku.ac.jp} Yurina Mizuno,$^1$
Tomoaki Matsumoto$^2$ \newauthor
and Kazuyuki Omukai$^{1}$\\
$^1$Astronomical Institute, Tohoku
University, Aoba, Sendai 980-8578, Japan \\ 
$^2$Faculty of Sustainability Studies, Hosei University, Fujimi, Chiyoda, Tokyo 102-8160, Japan
} 
\title[Fates of the dense cores formed by fragmentation]{Fates of the dense cores formed by fragmentation of filaments:
do they fragment again or not?}
\maketitle
\topmargin-1cm 

\begin{abstract}
Fragmentation of filaments into dense cores is thought to be an
important step in forming stars.  The bar-mode instability of
spherically collapsing cores found in previous linear analysis invokes a
possibility of re-fragmentation of the cores due to their ellipsoidal
(prolate or oblate) deformation.  To investigate this possibility,
here we perform three-dimensional self-gravitational hydrodynamics
simulations that follow all the way from filament fragmentation to
subsequent core collapse.  We assume the gas is polytropic with index
$\gamma$, which determines the stability of the bar-mode.
For the case that the fragmentation of isolated hydrostatic
filaments is triggered by the most unstable fragmentation mode, we find
the bar mode grows as collapse proceeds if $\gamma< 1.1$, in agreement
with the linear analysis.  However, it takes more than ten
orders-of-magnitude increase in the central density for the distortion
to become non-linear.
In addition to this fiducial case, we also study non-fiducial ones
such as the fragmentation is triggered by a fragmentation mode with a
longer wavelength and it occurs during radial collapse of filaments and
find the distortion rapidly grows.
In most of astrophysical applications, the effective polytropic index of
collapsing gas exceeds $1.1$ before ten orders-of-magnitude increase in
the central density.  Thus, supposing the fiducial case of filament
fragmentation, re-fragmentation of dense cores would not be likely and
their final mass would be determined when the filaments fragment.
\end{abstract}

\begin{keywords}
stars:formation\,--\,galaxies: star formation\,--\,galaxies: evolution.
\end{keywords}
 \section{Introduction}
 \label{sec:intro}

What determines the mass of stars?  In the standard scenario of star
formation, stars are formed inside dense cores with the mass conversion
efficiency of about $30$\%, as supported by both simulations
\citep[e.g.,][]{Machida:2012aa} and observations
\citep[e.g.,][]{Andre:2010aa}.
Dense cores, in turn, are thought to be formed by fragmentation of
filamentary molecular clouds, or simply filaments. While observations
with the {\it Herschel} satellite have revealed that the dense cores are
located along filaments \citep[e.g.,][]{Andre:2010aa,Arzoumanian:2011aa,Roy:2015aa}, simulations have
shown that filaments fragment into cores that subsequently
collapse in a run-away fashion \citep[e.g.,][]{Inutsuka:1997aa}.  At the
end of the collapse, which is well approximated with the self-similar
solution of spherical collapse \citep[so-called Larson-Penston
solution;][]{Larson:1969aa,Penston:1969ab,Yahil:1983aa}, protostars are
formed when the central parts of the dense cores become optically thick.

The theoretical works on filament fragmentation, both analytical
\citep[e.g.,][]{Stodolkiewicz:1963aa,Larson:1985aa,Nagasawa:1987aa,Inutsuka:1992aa,Fischera:2012aa}
and numerical \citep[e.g.,][]{Inutsuka:1997aa,
Heitsch:2013aa,Heitsch:2013ab,Heigl:2016aa,Clarke:2016ac,Gritschneder:2017aa}
ones, have shown that the typical fragmentation mass is given by the
Jeans mass at fragmentation of the filaments. This gives reasonable
estimate for the initial mass of dense cores, but the final mass can be
dramatically altered if the cores will fragment again in later
evolution.

Such re-fragmentation can be caused by deformation of the cores due to
the bar-mode instability of the Larson-Penston solution found in the
previous linear stability analyses \citep{Hanawa:2000aa,Lai:2000aa}. They have
shown that the  bar ($l=2$) mode is unstable if $\gamma<1.1$, where
$\gamma$ is the effective polytropic index of gas.  This instability
deforms cores into a prolate or oblate shape depending on the seed perturbation for the
instability. Unfortunately, the role of this instability in star
formation was unclear, because the linear analysis can predict
neither the initial amplitude of the unstable mode nor the fate of the
instability in the non-linear regime.  In order to address this issue,
it is necessary to perform numerical simulations.

In this paper, we investigate the possibility of re-fragmentation of
dense cores formed by filament fragmentation, focusing on the role of
the bar-mode instability.  We continuously follow filament fragmentation
and subsequent core collapse with three-dimensional self-gravitational
hydrodynamics simulations.  We study the $\gamma$ dependence of the
evolution of cores, as well as the dependence on the way in which
filaments fragment.  Although there have been a number of
simulations for core collapse after filament fragmentation
\citep[e.g.,][]{Nakamura:1993aa,Nakamura:2000aa,Matsumoto:1997aa,Inutsuka:1997aa},
none of them has focused on the possibility of re-fragmentation or the role of
the bar-mode instability.  

The paper is organized as follows. In Sec.~\ref{sec:method}, we describe
our models and numerical methods.  In Sec.~\ref{sec:results}, we
present the result of our simulations.  The conclusion and discussion
are given in Sec.~\ref{sec:conclusion}.

\section{Models \& Methods}
\label{sec:method}
 \subsection{Basics}
\label{sec:prop-inst}

Below, we briefly summarize the $\gamma$ dependence of the two types of
instabilities, namely the fragmentation-mode and bar-mode instabilities,
which motivates the models of this work.  We first introduce some useful
variables and then review the fragmentation-mode instability of
filaments and the bar-mode instability of the Larson-Penston solution.

The polytropic gas is characterized by the equation of state,
  \begin{align}
   P = K \rho^\gamma\,,
   \label{eq:1}
  \end{align}
and the sound speed is obtained as
  \begin{align}
   c_\mr{s}=\sqrt{K\gamma \rho^{\gamma-1}}\,.
   \label{eq:4}
  \end{align}
We define the free-fall time\footnote{For our convenience, we adopt the
definition of $t_\mr{ff}$ in equation~\eqref{eq:2}, instead of $(3/32\pi G
\rho)^{1/2}$, which is also used in the literature.}  as
  \begin{align}
   t_\mr{ff}=\frac{1}{\sqrt{4\pi G\rho}}\,,
   \label{eq:2}
  \end{align}
and the Jeans length as
  \begin{align}
   \lambda_\mr{J}=\sqrt{\frac{\pi}{G\rho}}c_\mr{s}=2\pi c_\mr{s}t_\mr{ff}\,.
   \label{eq:3}
  \end{align}
Once a reference density $\rho_0$ is given, we can obtain $c_\mr{s,0}$,
$t_\mr{ff,0}$ and $\lambda_\mr{J,0}$ for $\rho_0$ with
equations~\eqref{eq:4}\,--\,\eqref{eq:3}. Using these quantities and their
combinations, all dimensional quantities in this work can be made
dimensionless.

Infinitely long and static filaments are subject to gravitational instability of axisymmetric modes 
that leads to fragmentation into
cores.  The linear stability analysis of static polytropic filaments \citep[e.g.,][]{Larson:1985aa,Inutsuka:1992aa}
has shown that amplitude of the most unstable mode $\delta_\mr{F,max}$
grows exponentially with the growth rate $\sigma_\mr{F,max}$, as
\begin{align}
 \delta_\mr{F,max} \propto \exp\left[\sigma_\mr{F,max} \frac{t}{t_\mr{ff,0}}\right]\,.
 \label{eq:8}
\end{align}
We take the data for the wave number $k_\mr{max}$ and squared growth
rate $\mu_\mr{max}$ ($=\sigma_\mr{F,max}^2$) of the most unstable mode
from Fig.~9 of \cite{Inutsuka:1992aa} and fit them as functions of
$\gamma$ for $1\leq \gamma \leq 1.5$. As a result, we obtain the fitting
formulae,
\begin{align}
 k_\mr{max}H_0 =2.30-2.89\,\gamma+1.77\,\gamma^2-0.374\,\gamma^3\,,
 \label{eq:20}
\end{align}
and
\begin{align}
 \mu_\mr{max} =0.0688+0.236\,\gamma-0.268\,\gamma^2+0.0781\,\gamma^3\,,
 \label{eq:21}
\end{align}
where the central density of the filament is assumed to be $\rho_0$ and
$H_0=2c_\mr{s,0}/\sqrt{2\pi G
\rho_0}=(\sqrt{2}/\pi)\,\lambda_\mr{J,0}$ is the width of filaments
\citep[e.g.,][]{Inutsuka:1992aa}.  Table~\ref{tab:frag-inst} presents
the wavelength of the most unstable mode $\lambda_\mr{max}$
($=2\pi/k_\mr{max}$) and $\sigma_\mr{F,max}$ for selected values of
$\gamma$, computed with the above fitting formulae.

The linear analyses \citep[][]{Hanawa:2000aa,Lai:2000aa} have shown
the Larson-Penston solution is unstable if $\gamma<1.1$ due to the
bar-mode instability. The amplitude of the bar mode $\delta_{B}$ grows
in a power-law fashion with the growth rate $\sigma_\mr{B}$, as
\begin{align}
 \delta_{B} \propto \left(t_\mr{col}-t\right)^{-\sigma_\mr{B}}
 \propto \left(\rho_\mr{max}\right)^{\frac{\sigma_\mr{B}}{2}}\,.
 \label{eq:7}
\end{align}
To derive the second relation, we have used the relation
$\rho_\mr{max}\propto (t_\mr{col}-t)^{-2}$ for the Larson-Penston
solution, where $t_\mr{col}$ is the time when $\rho_\mr{max}$ formally
diverges.  Using the data for $\sigma_\mr{B}$ taken from Fig.~2 of
\cite{Hanawa:2000aa}, we obtain the fitting formula, 
 \begin{align}
  \sigma_\mr{B} = -2.84 + 9.39\,\gamma - 6.20\,\gamma^2\,,
 \end{align}
for $0.9\leq \gamma \leq 1.1$.  This formula is evaluated for some
values of $\gamma$ and shown in Table~\ref{tab:bar-inst}.\footnote{Note
that there is a room for numerical error even in the linear analysis
because the eigenmode is numerically obtained.  As a result, the
reported values of the critical $\gamma$ for the bar-mode instability,
$1.097$ in \cite{Hanawa:2000aa} and $1.09$ in \cite{Lai:2000aa}, are
similar but not exactly the same.  }

  \begin{table}
   \centering
   \caption{$\gamma$ dependence of the most unstable fragmentation mode}
   \label{tab:frag-inst}
   \begin{tabular}{lcccccc} \hline
    $\gamma$ & 0.9 & 0.95 & 1 & 1.05 & 1.1 & 1.2 \\\hline
    $\lambda_\mr{max}\,[H_{0}]\quad$ & 7.3 & 7.6 & 7.8 & 8.0 & 8.2 & 8.6\\
    $\sigma_\mr{F,max}\quad$ & 0.35 & 0.34 & 0.34 & 0.33 & 0.33 & 0.32\\\hline
   \end{tabular}
  \end{table}

  \begin{table}
   \centering
   \caption{$\gamma$ dependence of the bar mode}
   \label{tab:bar-inst}
   \begin{tabular}{lcccccc} \hline
    $\gamma$ & 0.9 & 0.95 & 1 & 1.05 & 1.1 & 1.2 \\\hline
    $\sigma_\mr{B}\quad$ & 0.59 & 0.48 & 0.35 & 0.18 & stable &stable \\\hline
   \end{tabular}
  \end{table}

\subsection{Models}
\label{sec:models}
We perform simulations that follow the fragmentation of filaments and
subsequent collapse of cores, assuming the gas is polytropic
(equation~\ref{eq:1}).  The initial conditions are generated by adding
velocity perturbations to filaments. Below, we will describe the models
studied in our simulations.

The initial density profile is given by
\begin{align}
\rho_\mr{ini}(r)=f\,\rho_\mr{st}(r)\,, \label{eq:10}
\end{align}
where $\rho_\mr{st}$ is the density profile of a hydrostatic polytropic
filament axisymmetric about the $z$ axis
\citep{Stodolkiewicz:1963aa,Ostriker:1964aa}, $r=\sqrt{x^2+y^2}$ the
cylindrical radius and $f$ a density enhancement factor.  Here, we take $\rho_0=\rho_\mr{st}(0)$.
For reference, the central density of filaments
in local star forming regions is about $10^{-20}-10^{-18}\cmr{g\,cm^{-3}}$
 \citep[see, e.g.,][]{Arzoumanian:2011aa}.
We numerically obtain $\rho_\mr{st}$ except for $\gamma=1$, when
$\rho_\mr{st}$ is obtained analytically as
$\rho_\mr{st}=\rho_0(1+r^2/H_0^2)^{-2}$ with
$H_0=2c_\mr{s,0}/\sqrt{2\pi G \rho_0}$.
We assume that the filament is under the external pressure of
an ambient gas with density $\rho_\mr{ext}$.

Fragmentation of filaments is triggered by the initial velocity
perturbations.  We assume a perturbation of $v_z$ with the sinusoidal
$z$ dependence and the initial velocity field is given by
\begin{align}
 \bm{v}_\mr{ini}(\bm{x})=  
 \left(\!\!\!
 \begin{array}{c}
  0\\
  0\\
  v_0\sin\left(2\pi z/\lambda\right)
 \end{array}
 \!\!\!
 \right)  
 \label{eq:11}\,,
\end{align}
where $v_0$ and  $\lambda$ are the amplitude and
wavelength of the perturbation, respectively.

The simulations are run with the model parameters summarized in
Table~\ref{tab:model}.  We first study the fiducial case where the
fragmentation of isolated hydrostatic filaments is triggered by the
most unstable mode, as often supposed in the literature
\citep[e.g.,][]{Larson:1985aa}.  To investigate the $\gamma$ dependence
of the evolution, we perform the simulations of set ``G'' (for
``gamma'') with different $\gamma$ ($=0.9$, $0.95$, $1$, $1.05$, $1.1$
and $1.2$).  In this case, we take $f=1$,
$v_0=10^{-3}\,c_\mr{s,0}$, $\lambda=\lambda_\mr{max}$ and
$\rho_\mr{ext}=0$.  (see Sec.~\ref{sec:prop-inst}).  The very small
perturbation is adopted just to follow the fragmentation triggered
purely by the most unstable eigenmode and the evolution is
astrophysically relevant only after the mode grows up to have a certain
amplitude.

In the actual astrophysical situations, however, fragmentation can
proceed in a non-fiducial way.  Thus we perform four sets of simulations
with the model parameters different from the fiducial ones.  The
parameters are set to the fiducial values unless otherwise stated.
First, we study the effect of the deviation of the initial velocity
perturbation from the most unstable eigenmode with the simulations of
set ``V'' (for ``velocity''). For this set, we enhance $v_0$ to $0.1$,
$0.3$ and $0.5\,c_\mr{s,0}$ (see equation~\ref{eq:11}) and emphasize the
converging nature of the initial velocity field.  Second, we study
fragmentation triggered by modes with various wavelength by performing
the simulations of set ``L'' (for ``lambda''), for which we change
$\lambda$ to $0.6$, $1.5$ and $2\,\lambda_\mr{max}$ (see
equation~\ref{eq:11}).  Third, to study fragmentation of radially
collapsing filaments, we perform the simulations of set ``D'' (for
``density''), for which we enhances the initial density taking $f$ as
$1.05$, $1.1$ and $1.2$ (see equation~\ref{eq:10}), with $v_0$ also
increased to $0.1$ and $0.5\,c_\mr{s,0}$.  In the above three cases, we
take $\gamma = 1$, $1.05$ and $1.2$ to see the $\gamma$ dependence.
Finally, to study fragmentation of filaments under external pressure of
an ambient gas, we perform the simulations of set ``E'' (for
``external''), for which we change the density of an ambient gas by
taking $\rho_\mr{ext} = 0.01$, $0.04$ and $0.09$. We take $\gamma = 1$
for set E.

  \begin{table}
   \centering
   \caption{Model parameters for runs studied}
   \label{tab:model}
   \begin{tabular}{lcccccc} \hline
    Set&$\gamma$&$\lambda/\lambda_\mr{max}$&$f$&$v_0/c_\mr{s,\,0}$&$\rho_\mr{ext}/\rho_0$$^{a}$&shape$^{b}$ \\\hline
    &{\boldmath$0.9$}&$1$&$1$&$10^{-3}$&0&P\\
    &{\boldmath$0.95$}&$1$&$1$&$10^{-3}$&0&P\\
    G&{\boldmath$1$}&$1$&$1$&$10^{-3}$&0&O\\
    &{\boldmath$1.05$}&$1$&$1$&$10^{-3}$&0&O\\
    &{\boldmath$1.1$}&$1$&$1$&$10^{-3}$&0&S\\
    &{\boldmath$1.2$}&$1$&$1$&$10^{-3}$&0&S\\\hline

    &$1$&$1$&$1$&{\boldmath$0.1$}&0&O\\
    &$1$&$1$&$1$&{\boldmath$0.3$}&0&O\\
    &$1$&$1$&$1$&{\boldmath$0.5$}&0&O\\
    &$1.05$&$1$&$1$&{\boldmath$0.1$}&0&O\\
    V&$1.05$&$1$&$1$&{\boldmath$0.3$}&0&O\\
    &$1.05$&$1$&$1$&{\boldmath$0.5$}&0&O\\
    &$1.2$&$1$&$1$&{\boldmath$0.1$}&0&S\\
    &$1.2$&$1$&$1$&{\boldmath$0.3$}&0&S\\
    &$1.2$&$1$&$1$&{\boldmath$0.5$}&0&S\\\hline

    &$1$&{\boldmath$0.6$}&$1$&$10^{-3}$&0&O\\
    &$1$&{\boldmath$1.5$}&$1$&$10^{-3}$&0&P\\
    &$1$&{\boldmath$2$}&$1$&$10^{-3}$&0&P\\
    &$1.05$&{\boldmath$0.6$}&$1$&$10^{-3}$&0&O\\
    L&$1.05$&{\boldmath$1.5$}&$1$&$10^{-3}$&0&P\\
    &$1.05$&{\boldmath$2$}&$1$&$10^{-3}$&0&P\\
    &$1.2$&{\boldmath$0.6$}&$1$&$10^{-3}$&0&S\\
    &$1.2$&{\boldmath$1.5$}&$1$&$10^{-3}$&0&S\\
    &$1.2$&{\boldmath$2$}&$1$&$10^{-3}$&0&S\\\hline

    &$1$&$1$&{\boldmath$1.05$}&$0.1$&0&P\\
    &$1$&$1$&{\boldmath$1.1$}&$0.1$&0&P\\
    &$1$&$1$&{\boldmath$1.2$}&$0.1$&0&P\\
    &$1.05$&$1$&{\boldmath$1.05$}&$0.1$&0&P\\
    &$1.05$&$1$&{\boldmath$1.1$}&$0.1$&0&P\\
    &$1.05$&$1$&{\boldmath$1.2$}&$0.1$&0&P\\
    &$1.2$&$1$&{\boldmath$1.05$}&$0.1$&0&S\\
    &$1.2$&$1$&{\boldmath$1.1$}&$0.1$&0&S\\
    D&$1.2$&$1$&{\boldmath$1.2$}&$0.1$&0&S\\
    &$1$&$1$&{\boldmath$1.05$}&$0.5$&0&O\\
    &$1$&$1$&{\boldmath$1.1$}&$0.5$&0&P\\
    &$1$&$1$&{\boldmath$1.2$}&$0.5$&0&P\\
    &$1.05$&$1$&{\boldmath$1.05$}&$0.5$&0&O\\
    &$1.05$&$1$&{\boldmath$1.1$}&$0.5$&0&O\\
    &$1.05$&$1$&{\boldmath$1.2$}&$0.5$&0&P\\
    &$1.2$&$1$&{\boldmath$1.05$}&$0.5$&0&S\\
    &$1.2$&$1$&{\boldmath$1.1$}&$0.5$&0&S\\
    &$1.2$&$1$&{\boldmath$1.2$}&$0.5$&0&S\\\hline

    &$1$&$1$&$1$&$0.1$&{\boldmath $0.01$}&O\\
    E&$1$&$1$&$1$&$0.1$&{\boldmath $0.04$}&O\\
    &$1$&$1$&$1$&$0.1$&{\boldmath $0.09$}&O\\\hline
   \end{tabular}\\
   \begin{flushleft}
    $^{a}$The density at the boundary of the computational box is finite
    even when $\rho_\mr{ext}/\rho_0=0$ for computational reason (see
    text).\\ $^{b}$Shape of core at the end of simulation: ``P'', ``O''
    and ``S'' indicate prolate, oblate and spherical shapes,
    respectively.
   \end{flushleft}
  \end{table}

  \subsection{Numerical methods}
  \label{sec:numerical}

We use the self-gravitational magneto-hydrodynamics code with adaptive
mesh refinement (AMR), {\tt SFUMATO} \citep{Matsumoto:2007aa}, but with
the magnetic module switched off.  The hydrodynamical solver adopts the
total variation diminishing cell-centred scheme with second-order
accuracy in space and time.

Our computational domain is a cube with the side length
$L_\mr{box}=\lambda$ (see equation~\ref{eq:11}).  We solve the
three-dimensional (3D) hydrodynamics with the Cartesian coordinate set,
without using the axisymmetry of the system.  We initially set a uniform
grid with $N_\mr{ini}=256$ meshes in each direction ($256^3$ cells).
The Jeans condition is employed as a refinement criterion in the
block-structured AMR technique of the code. Blocks are refined to
resolve one Jeans length $\lambda_\mr{J}$ (equation~\ref{eq:3}) with at
least $N_\mr{ref}=32$ meshes.  We test the convergence of the numerical results in
Appendix~\ref{sec:resolution_dependence}.

We assume the periodic boundary condition in the $z$ direction.  In the
$r$ direction, however, we fix the density to the boundary value
$\rho_\mr{b}$ and the velocity to zero outside the boundary at
$r_\mr{b}$.  We take $r_\mr{b}$ as the radius where the initial density
is equal to the ambient density, i.e.,
$\rho_\mr{ini}(r_\mr{b})=\rho_\mr{ext}$, and $\rho_\mr{b}$ as
$\rho_\mr{ext}$.  If the cylinder with $r_\mr{b}$ is larger than the
computational box, $r_\mr{b}$ and $\rho_\mr{b}$ are replaced with
$L_\mr{box}/2$ and $\rho_\mr{ini}(L_\mr{box}/2)$, respectively.  For the
case $\rho_\mr{b}=0$, we take a finite but sufficiently small value
(e.g., $\rho_\mr{b}=10^{-8}\rho_0$) in the actual calculations for
computational reason. We assume the gravitational potential for the
isolated filament on the surfaces of the computational box in the $x$
and $y$ directions.

  \section{Results}
  \label{sec:results}

  \subsection{Evolution in a typical case}
  \label{sec:time_evolution}

  \begin{figure*}
   \centering \includegraphics[width=13.5cm]{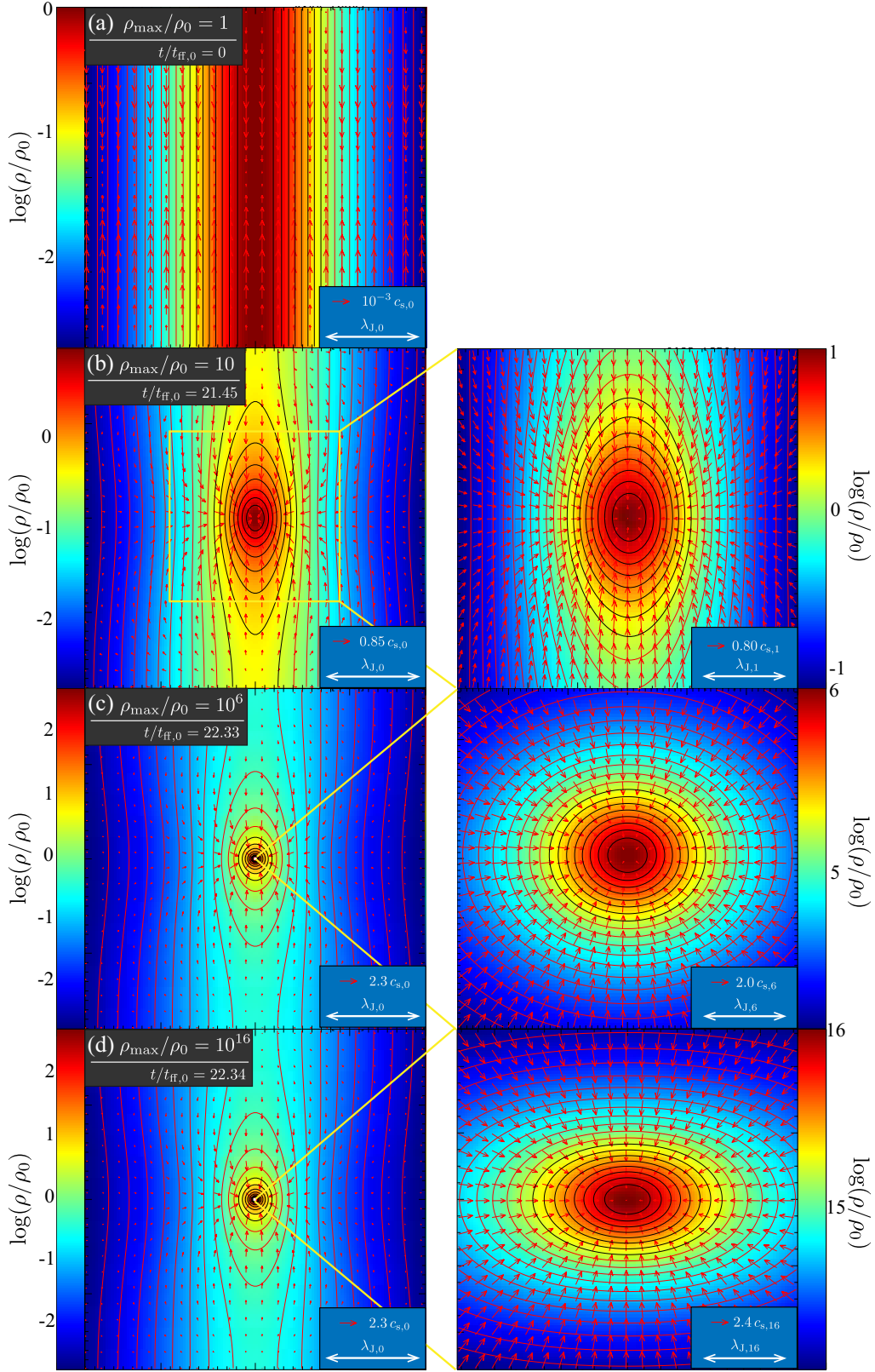}
   \caption{Time evolution of density (colour) and velocity (arrows)
   distributions in the $xz$-plane for the $\gamma=1.05$ model of the
   fiducial case (set G in Table~\ref{tab:model}).  The whole
   computational domains (left) and the central boxes on a Jeans
   length scale (right) are shown.  The maximum (central) density is (a)
   $\rho_\mr{max}/\rho_0=1$ (initial time), (b) $10$, (c) $10^6$ and (d)
   $10^{16}$.  The velocity and length scales are shown on
   the bottom-right corner of each panel, with $c_\mr{s,n} =
   10^{n(\gamma-1)/2}\, c_\mr{s,0}$ (equation~\ref{eq:4}) and $\lambda_\mr{J,n} =
   10^{n(\gamma-2)/2}\,\lambda_\mr{J,0}$ (equation~\ref{eq:3}) for $\rho = 10^n\,
   \rho_0$.}
   \label{fig:sanpshots_g105}
  \end{figure*}

  \begin{figure}
   \centering \includegraphics[width=8cm]{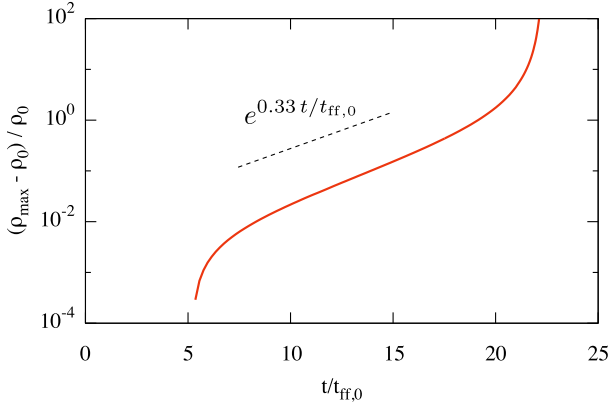}
   \caption{Time evolution of $\delta = (\rho_\mr{max}-\rho_0)/\rho_0$
   for the model in Fig.~\ref{fig:sanpshots_g105}.
   The dashed line represents the
   linear growth rate (see equation~\ref{eq:8}).  } \label{fig:rhomax_g105}
  \end{figure}

  \begin{figure}
   \centering
   \includegraphics[width=8cm]{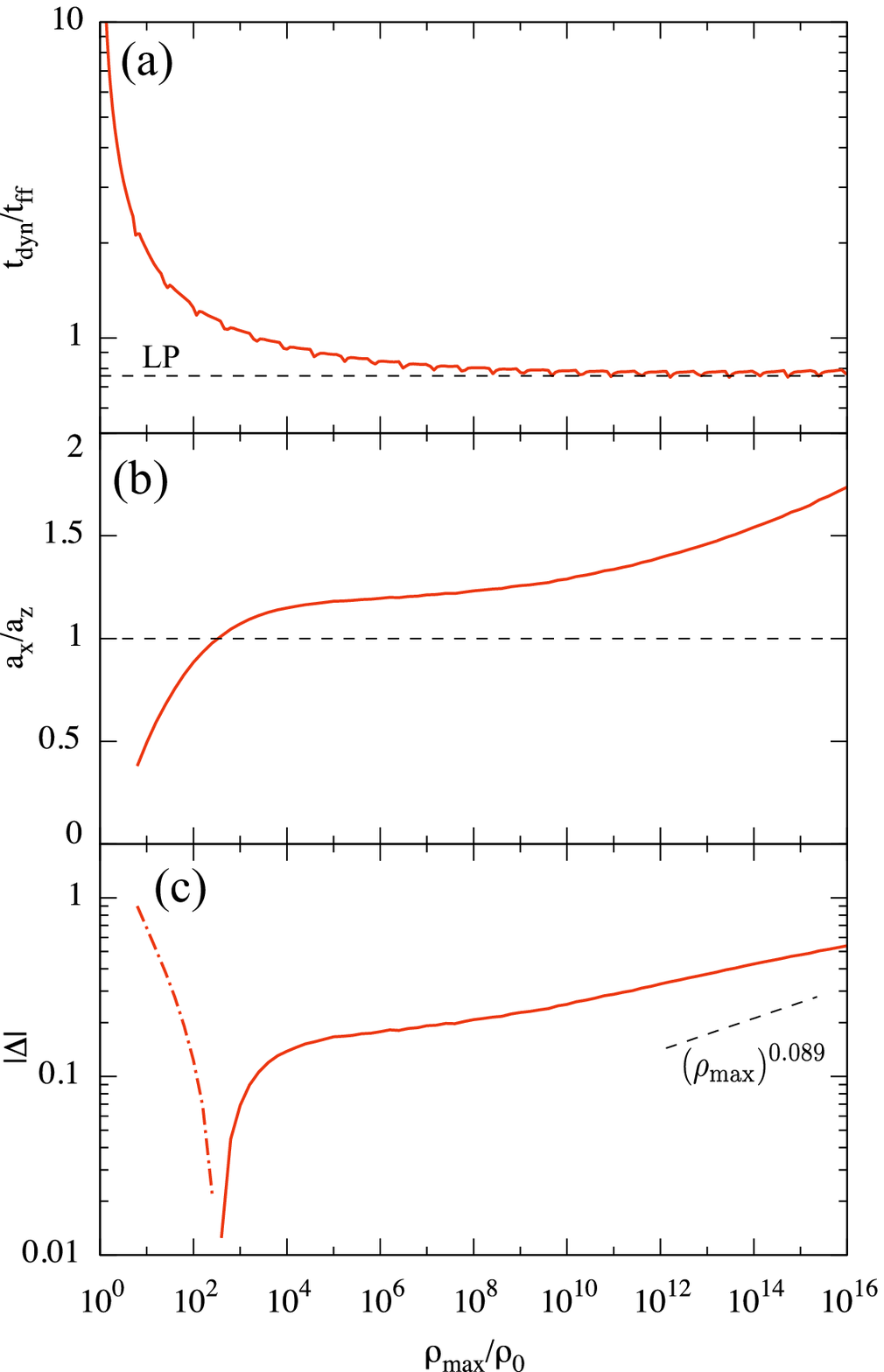}
   \caption{Time evolution of (a) $t_\mr{dyn}/t_\mr{ff}$, (b) $a_x/a_z$
   and (c) $\Delta = 2(a_x-a_z)/(a_x+a_z)$ for the model in
   Fig.~\ref{fig:sanpshots_g105}.   In panel (a), the dashed line represents
   $t_\mr{dyn}/t_\mr{ff}$ of the Larson-Penston solution for
   $\gamma=1.05$.  In panel (c), the solid (dot-dashed) line corresponds
   to a positive (negative) value, while the dashed line represents the
   linear growth rate (see equation~\ref{eq:7}).  }
   \label{fig:tdynff_rat_oblt_g105}
  \end{figure}

In this section, we describe the time evolution in the $\gamma=1.05$
model of the fiducial case (fourth line in Table~\ref{tab:model}), as a
typical example of our simulations.  Although the shapes of the cores at
the end of simulations are greatly different depending on the models,
the evolution generally proceeds in a way similar to that described
below.

Fig.~\ref{fig:sanpshots_g105} shows the density and velocity
distributions in the $xz$-plane at the four different stages of
evolution: (a) the maximum (central) density $\rho_\mr{max}/\rho_0=1$
(initial time), (b) $10$, (c) $10^6$ and (d) $10^{16}$. Note that we use
$\rho_\mr{max}/\rho_0$ as a time variable since it monotonically
increases with the time as collapse proceeds.  The panels in
Fig.~\ref{fig:sanpshots_g105} show that filament fragmentation and
subsequent collapse of the core proceed as follows:
\begin{enumerate}
 \item[(a)] A small velocity perturbation to the filament is seen in the
	    initial condition. It is a seed for the most unstable
	    fragmentation mode, which later grows and leads to the
	    fragmentation of the filament.

 \item[(b)] Left: A high density region, or core, is formed
	    due to the fragmentation of the filament and starts
	    collecting the surrounding gas gravitationally. Right: The
	    core is initially prolate along the filament.

 \item[(c)] Left: The core collapses in a run-away fashion.
	    Right: The gas dynamics near the centre approaches the
	    Larson-Penston solution. The formerly prolate core becomes
	    nearly spherical, although slightly oblate.

 \item[(d)] Left: The run-away collapse continues.  Right: Once the gas
	    dynamics becomes sufficiently close to the Larson-Penston
	    solution, the bar-mode instability begins to
	    grow. Accordingly, the core becomes more and more oblate and
	    the distortion finally becomes non-linear.
\end{enumerate}

Below, we will examine the evolution in more detail, focusing on the
fragmentation of the filament and the collapse and deformation of the core.

First, we see how fragmentation occurs.  To quantify the degree of
fragmentation, we define the overdensity $\delta$ normalized by
$\rho_0=\rho_\mr{ini}(0)$ (equation~\ref{eq:10}) as
\begin{align}
 \delta = \frac{\rho_\mr{max}(t)-\rho_0}{\rho_0}\,,
 \label{eq:5}
\end{align}
where $\rho_\mr{max}(t)$ is the maximum density at $t$. The density is
largest at the centre of the core since the beginning of the
fragmentation.  The fragmentation is roughly completed when $\delta \sim
O(1)$.  Fig.~\ref{fig:rhomax_g105} shows the time evolution of $\delta$,
along with the linear growth rate of the most unstable fragmentation
mode ($\sigma_\mr{F,max}=0.33$; Table~\ref{tab:frag-inst}).  The
agreement of the two slopes for $10^{-2}\lesssim \delta \lesssim 1$
indicates that the fragmentation is caused by the growth of the most
unstable mode, as we expect for the fiducial case. Note that only the result for
$\delta\gtrsim 10^{-2}$ is reliable because that for lower $\delta$
depends on the resolution (see
Appendix~\ref{sec:resolution_dependence}).

Second, we examine the convergence of the core collapse to the
Larson-Penston solution.  To quantify how much the central dynamics is
close to the Larson-Penston solution, we introduce the ratio
$t_\mr{dyn}/t_\mr{ff}$,\footnote{ For the same purpose,
\cite{Tsuribe:1999aa} introduced the normalized central density $z_0$
using a relation $\rho_\mr{max}=z_0/(t_\mr{col}-t)^2$.  We can show
$t_\mr{dyn}/t_\mr{ff}=\sqrt{z_0}/2$ with
equation~\eqref{eq:9}.}  where the dynamical time scale
$t_\mr{dyn}$ is defined as
\begin{align}
 t_\mr{dyn}= \frac{\rho_\mr{max}}{\dot{\rho}_\mr{max}}\,,
\label{eq:9}
\end{align}
and the free-fall time scale $t_\mr{ff}$ is given by
equation~\eqref{eq:2} with $\rho=\rho_\mr{max}$.  This ratio indicates
the rapidness of the collapse: $t_\mr{dyn}/t_\mr{ff}=0.41$ for the
homogeneous gravitational collapse \citep[see, e.g.,][]{Tsuribe:1999aa}
and it increases to $t_\mr{dyn}/t_\mr{ff}=0.76$ for the Larson-Penston
solution with $\gamma=1.05$ \citep[see, e.g.,][]{Lai:2000aa}, because
the pressure delays the collapse.  The evolution of
$t_\mr{dyn}/t_\mr{ff}$ is shown in
Fig.~\ref{fig:tdynff_rat_oblt_g105}(a). Soon after the fragmentation,
the collapse is still slow and $t_\mr{dyn}/t_\mr{ff}$ is large. Later
on, the dynamics of core approaches the Larson-Penston solution
\citep[e.g.,][]{Inutsuka:1997aa}, as indicated by the decrease of
$t_\mr{dyn}/t_\mr{ff}$ toward the value for the Larson-Penston solution
(horizontal dashed line). However, the approach is rather slow and it
takes about five orders of magnitude in density increase for the collapse to become
close to the Larson-Penston solution. Note that the features of
$t_\mr{dyn}/t_\mr{ff}$ appearing every four-time increase in the density
are originated from numerical errors at refinement, although they hardly
affect the result (Appendix~\ref{sec:resolution_dependence}).

Finally, we see how the core deforms. We use the
axial ratio to quantify the deformation, where the axes of the core are
defined using the inertia tensor \citep[see, e.g.,][]{Matsumoto:1999aa},
as below.  We regard the central dense region with
$\rho > \rho_\mr{th}=0.1\,\rho_\mr{max}$ as the core, 
and then its inertia tensor and total mass are given respectively by
\begin{align}
 I_{ij}=\int_{\rho>\rho_\mr{th}} \mr{d}\bm{x}\, x^ix^j \rho(\bm{x})\,,
 \label{eq:16}
\end{align}
and 
\begin{align}
 M=\int_{\rho>\rho_\mr{th}} \mr{d}\bm{x}\,
 \rho(\bm{x})\,.
\label{eq:12}
\end{align}
We can assume that the three eigenvalues of $I_{ij}/M$, denoted as
$\lambda_i$ ($i=1$, $2$ and $3$),  
satisfy $\lambda_1=\lambda_2$ without loss of generality thanks to the
axisymmetry of the core.
Thus, we define the axes of the core in the $x$, $y$ and $z$ directions as
\begin{align}
 a_x = a_y = \sqrt{\lambda_1}\,,\quad
 a_z =\sqrt{\lambda_3}\,,
 \label{eq:6}
\end{align}
respectively.  We have checked that adopting different
$\rho_\mr{th}$ has little influence on our results.

 \begin{figure*}
  \centering \includegraphics[width=16cm]{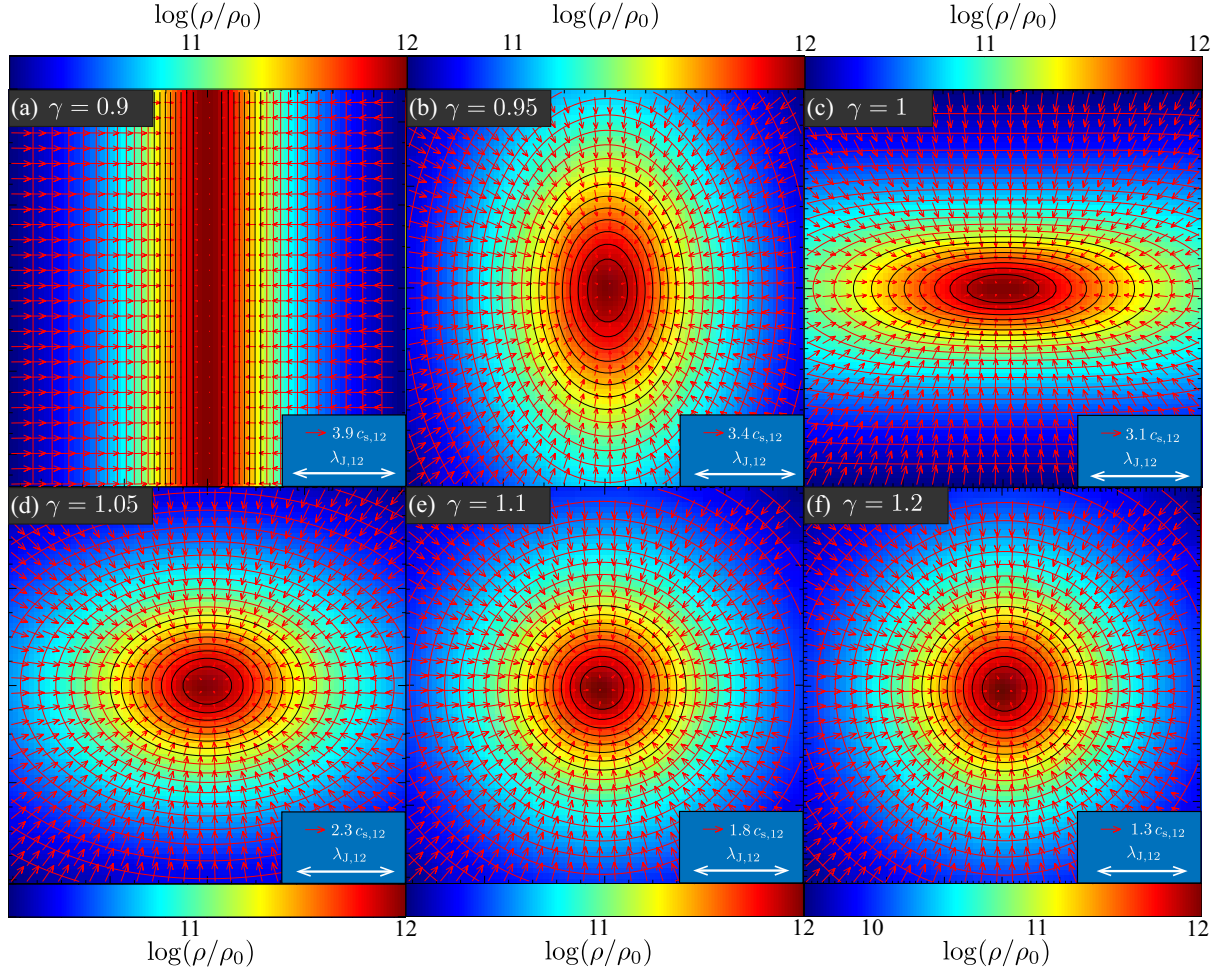}
  \caption{Same as the right column of Fig.~\ref{fig:sanpshots_g105} but for the models of
  the fiducial case with (a) $\gamma=0.9$, (b) $0.95$, (c) $1$, (d)
  $1.05$, (e) $1.1$ and (f) $1.2$.  The maximum (central) density is
  $\rho_\mr{max}/\rho_0=10^{12}$ in all panels.   
  }
  \label{fig:snapshots_gamma}
 \end{figure*}

The evolution of the axial ratio $a_x/a_z$ ($=a_y/a_z$) of the core is
shown in Fig.~\ref{fig:tdynff_rat_oblt_g105}(b), where $a_x/a_z <1$,
$=1$ and $>1$ correspond to prolate, spherical and oblate shapes,
respectively.  Note the definition of the axes of the core is not valid and thus
we do not plot $a_x/a_z$ for $\rho_\mr{max}/\rho_0 \lesssim 10$ because
the integration region in equations~\eqref{eq:16} and \eqref{eq:12} is
not confined in the computational domain.  As expected, the evolution of
$a_x/a_z$ is consistent with the density profiles seen in
Fig.~\ref{fig:sanpshots_g105}.  We see that $a_x/a_z$ is initially small
but rapidly increases and eventually exceeds unity. Afterwards, it stops
increasing and remains almost constant with $a_x/a_z\gtrsim 1$ for a
while, although it finally begins to increase again due to the bar-mode
instability.

To analyze the growth of the bar mode in more detail, we define the
oblateness \citep[see][]{Matsumoto:1999aa} as
\begin{align}
 \Delta = \frac{a_x-a_z}{(a_x+a_z)/2}\,,
 \label{eq:19}
\end{align}
where $a_x$ ($=a_y$) and $a_z$ are defined in equation~\eqref{eq:6}.
Fig.~\ref{fig:tdynff_rat_oblt_g105}(c) shows the evolution of $\Delta$.
The oblateness $\Delta$ changes its sign at $\rho_\mr{max}/\rho_0\sim
10^2$, as the axial ratio changes from $a_x/a_z<1$ to $a_x/a_z>1$
(Fig.~\ref{fig:tdynff_rat_oblt_g105}b), and begins to increase in a
power-law fashion with a small initial amplitude of $\Delta\lesssim 0.1$
at $\rho_\mr{max}/\rho_0\sim 10^5$, when the collapse becomes
sufficiently close to the Larson-Penston solution
(Fig.~\ref{fig:tdynff_rat_oblt_g105}a).  The rate of the power-law
increase is consistent with the linear analysis of the bar-mode
instability ($\sigma_\mr{B}=0.089$; Table~\ref{tab:bar-inst}),\footnote{
It can be shown that the amplitude of the bar mode $\delta_\mr{B}$
(equation~\ref{eq:7}) is proportional to the oblateness $\Delta$
(equation~\ref{eq:19}) in the linear regime.}  indicating that the
distortion is caused by the bar mode. We emphasize here that the bar
mode grows only when the collapse is sufficiently close to the
Larson-Penston solution.

The growth rate is slightly smaller than in the linear analysis due to
several reasons. First, the background collapse is not exactly the
Larson-Penston solution. Second, not only the pure bar mode but also
other modes contribute to $\Delta$. Third, the growth rate tends to be
smaller in the non-linear regime owing to the small dynamic range of
$\Delta$ ($-2<\Delta<2$).  Note also that the growth rate can be
overestimated in the linear analysis \citep{Hanawa:2000aa,Lai:2000aa}
due to numerical error (see the footnote in Sec.~\ref{sec:prop-inst}).

\subsection{$\gamma$ dependence}
\label{sec:gamma}

  \begin{figure}
   \centering \hspace*{-0.2cm}
   \includegraphics[width=8.5cm]{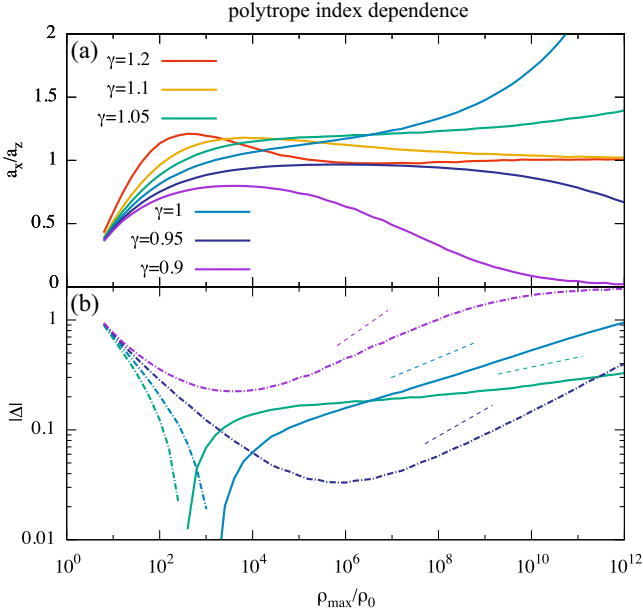} \caption{Panel
   (a): same as Fig.~\ref{fig:tdynff_rat_oblt_g105}(b) but for the
   models in Fig.~\ref{fig:snapshots_gamma} with $\gamma=1.2$ (red), $1.1$ (orange), $1.05$ (green), $1$
   (blue), $0.95$ (dark blue) and $0.9$ (purple).
   Panel (b): same as Fig.~\ref{fig:tdynff_rat_oblt_g105}(c) but for the above models with $\gamma\le 1.05$.
    The dashed lines are the linear growth rates for
   corresponding $\gamma$ (see equation~\ref{eq:8}).  }
   \label{fig:rat_oblt_gamma}
  \end{figure}

Here we investigate the $\gamma$ dependence of the evolution in the
fiducial case, where fragmentation of isolated hydrostatic filaments is triggered
by the most unstable mode, as often considered in the literature
\citep[e.g.,][]{Larson:1985aa}.  Below, we present the results for the
models with $\gamma=0.9$, $0.95$, $1$, $1.05$, $1.1$ and $1.2$ (set G in
Table~\ref{tab:model}).

Fig.~\ref{fig:snapshots_gamma} shows the final shapes of the cores at
$\rho_\mr{max}/\rho_0=10^{12}$.  Although the evolution generally
proceeds in a way similar to that explained in
Sec.~\ref{sec:time_evolution} irrespective of $\gamma$, the final shapes
are greatly different depending on $\gamma$, as summarized in
Table~\ref{tab:shape}.  Below, we will see how this dependence arises.

  \begin{table}
   \centering
   \caption{Summary for the shapes of the cores in Fig.~\ref{fig:snapshots_gamma}}
   \label{tab:shape}
   \begin{tabular}{lc} \hline
\hspace*{0.5cm}$\gamma$\hspace{2cm} &\hspace*{1.2cm} Shape \hspace*{1.2cm}\\\hline
\hspace*{0.3cm}(a) $0.9$ & strongly prolate\\
\hspace*{0.3cm}(b) $0.95$ & weakly prolate\\
\hspace*{0.3cm}(c) $1$ & moderately oblate\\
\hspace*{0.3cm}(d) $1.05$ & weakly oblate\\
\hspace*{0.3cm}(e) $1.1$  & spherically symmetric\\
\hspace*{0.3cm}(f) $1.2$ & spherically symmetric\\\hline
   \end{tabular}
  \end{table}

Fig.~\ref{fig:rat_oblt_gamma}(a) shows the evolution of the axial ratio
$a_x/a_z$. Although $a_x/a_z$ generally increases in the initial phase
($\rho_\mr{max}/\rho_0\lesssim10^{2}$), it evolves differently later on
depending on the value of $\gamma$.  In the $\gamma= 0.9$ model, $a_x/a_z$ begins to
decrease before reaching unity due to the rapid growth of bar-mode instability,
and thus the core becomes prolate. In the less unstable case with
$\gamma= 0.95$, the evolution is similar but the final distortion is
weaker.  In the models with $\gamma=1$ and $1.05$, the bar-mode
instability is so weak that $a_x/a_z$ exceeds unity before the
instability begin to grow and thus the cores become oblate.  The
oblateness is stronger in the model with $\gamma=1$ than with
$\gamma=1.05$ because the growth rate of the bar-mode instability is
larger for smaller $\gamma$.  In the models with $\gamma=1.1$ and $1.2$,
the cores become spherical because the spherical collapse is stable and
the bar-mode perturbation damps.

How the growth of the bar-mode instability depends on $\gamma$ is
clearly seen in Fig.~\ref{fig:rat_oblt_gamma} (b), where we plot the
evolution of the oblateness $\Delta$ (equation~\ref{eq:19}) for the
models with $\gamma\le 1.05$.  Except for the $\gamma=0.9$ model, the
bar mode begins to grow in a power-law fashion with a small initial
amplitude of $\Delta\lesssim 0.1$ at $\rho_\mr{max}/\rho_0 \sim
10^5\,\text{--}\,10^6$, when the background spherical collapse becomes
close to the Larson-Penston solution, as seen for the $\gamma=1.05$ case
in Sec.~\ref{sec:time_evolution}.  Because of this, in astrophysically interesting
cases of $1\leq \gamma < 1.1$ (see Sec.~\ref{sec:conclusion}), it needs about ten orders-of-magnitude
increase in the density before the distortion becomes significant.  In
the $\gamma=0.9$ model, the bar mode is so strong that it begins to grow
in the relatively early stage of the evolution ($\rho_\mr{max}/\rho_0
\sim 10^4$) with a certain initial amplitude ($\Delta\gtrsim 0.1$).  It
is also seen that the bar mode grows faster for smaller $\gamma$, as
expected from the linear analysis \citep{Hanawa:2000aa,Lai:2000aa}.  The
growth rates agree well with those obtained by the linear
analysis (dashed lines), confirming that the distortion is caused by
the bar-mode instability.

In summary, the $\gamma$ dependence of the final shapes of the cores
can be understood from the $\gamma$ dependence
of the bar-mode instability.  The core becomes spherical if
the bar mode is stable ($\gamma\ge1.1$), whereas it tends to deform
if the bar mode is unstable ($\gamma<1.1$).  In the case with strong
instability ($\gamma<1$), the deformation begins before the axial ratio
reaches unity and the core becomes prolate.  In the case
with weak instability ($1\le\gamma<1.1$), however, the deformation begins
after the axial ratio exceeds unity and the core becomes oblate.

  \subsection{Non-fiducial cases}
  \label{sec:dep-initial}

Here, we investigate the fragmentation of filaments proceeding in a
non-fiducial way, i.e., the cases different from the fragmentation of
isolated hydrostatic filaments triggered by the most unstable
fragmentation mode. Astrophysically, various situations can be
encountered: for example, unstable modes other than the most unstable
one can trigger fragmentation, fragmentation can occur during radial
collapse of filaments, filaments are not isolated but under the external
pressure of an ambient gas, etc.

Considering these possibilities, we perform simulations with the
different amplitude and wavelength of the initial velocity perturbation
in Sec.~\ref{sec:amplitude} and Sec.~\ref{sec:wavelength}, respectively.
We then perform simulations with the enhanced filament density in
Sec.~\ref{sec:density}.  To see the $\gamma$ dependence, we take
$\gamma=1$, $1.05$ and $1.2$ for the cases above. Finally, we study
fragmentation of filaments under external pressure in
Sec.~\ref{sec:p_ext}.  The parameters are fixed to the fiducial values,
unless otherwise stated.

As will be shown below, the results in this section suggest that
the evolution of the cores can be largely affected by how filament
fragmentation proceeds.  Thus, the results obtained for the fiducial
case should be treated with caution in astrophysical applications.

  \subsubsection{Fragmentation by perturbation with large amplitude}
  \label{sec:amplitude}
  \begin{figure}
   \centering \includegraphics[width=8cm]{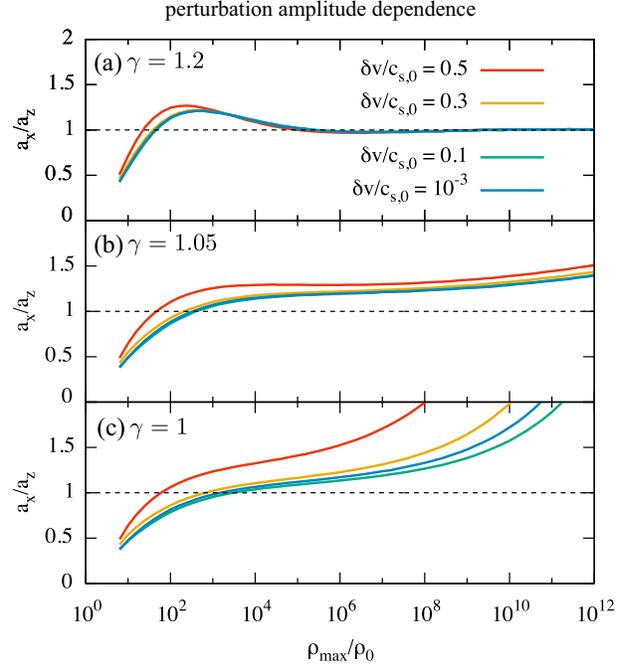}
   \caption{Same as Fig.~\ref{fig:rat_oblt_gamma}(a) but for the models
   studied in Sec.~\ref{sec:amplitude} with (a) $\gamma=1.2$, (b) $1.05$ and (c) $1$.  The
   amplitude of initial velocity perturbation $v_0/c_\mr{s,0}$ is $0.5$
   (red), $0.3$ (orange), $0.1$ (green) and $10^{-3}$ (blue).  The lines
   for $v_0/c_\mr{s,0}=0.3$, $0.1$ and $10^{-3}$ in panel (a) and those
   for $v_0/c_\mr{s,0}=0.1$ and $10^{-3}$ in panel (b) are overlapped
   and cannot be separately seen.  } \label{fig:ratio_velocity}
  \end{figure}

Here, we present the results of our simulations for the models with
different amplitudes of the initial velocity perturbation.  Since the
initial perturbation is not exactly the most unstable eigenmode, the
converging nature of the initial velocity field becomes important as the
initial amplitude increases.  The amplitude is taken to be
$v_0/c_\mr{s,0}=10^{-3}$ (fiducial), $0.1$, $0.3$ and $0.5$ (set V and a part of set G in Table~\ref{tab:model}).

Fig.~\ref{fig:ratio_velocity} shows the evolution of $a_x/a_z$ for (a)
$\gamma=1.2$, (b) $1.05$ and (c) $1$.  For all $\gamma$, the dependence
on the initial amplitude is small unless $\delta v/c_\mr{s,0}$ is as
large as $0.5$. In the case $\delta v/c_\mr{s,0}=0.5$, the core is
compressed due to the converging initial velocity field
(equation~\ref{eq:11}) and becomes more oblate than in the other cases
in the early phase ($\rho_\mr{max}/\rho_0 \lesssim 10^2$).  The
subsequent evolution is similar to the other cases for (a) $\gamma=1.2$
and (b) $1.05$. For (c) $\gamma=1$, however, the distortion becomes
non-linear at somewhat smaller $\rho_\mr{max}/\rho_0$ due to the larger
oblateness in the early phase.

These results suggest that the evolution is almost the same as the fiducial
case, where the most unstable eigenmode triggers the fragmentation, as
long as $\delta v/c_\mr{s,0}\leq 0.3$. This is because the most unstable
mode grows and dominates the other modes before fragmentation, although
the initial perturbation given by equation~\eqref{eq:11} is not exactly
the most unstable eigenmode.  We conclude that the dependence on the
amplitude of the initial velocity perturbation is weak as long as $\delta
v/c_\mr{s,0}\lesssim 0.3$.

  \subsubsection{Fragmentation by perturbation with different wavelength}
  \label{sec:wavelength}

  \begin{figure}
   \centering \includegraphics[width=8cm]{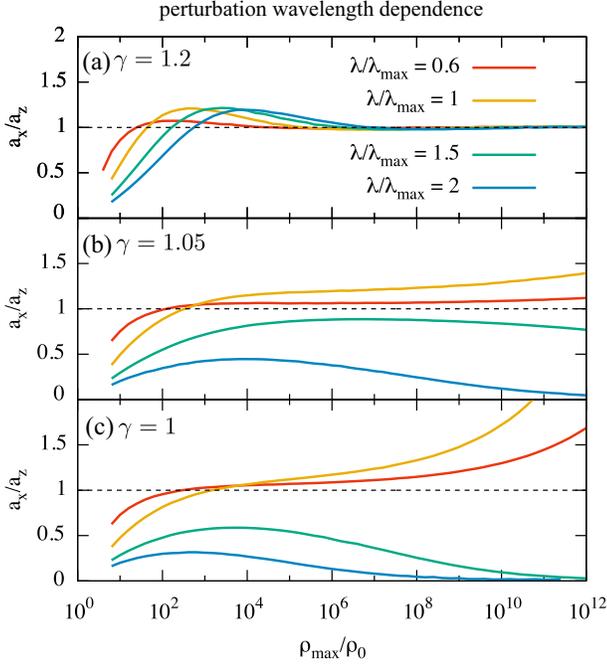}
   \caption{Same as Fig.~\ref{fig:rat_oblt_gamma}(a) but for the models
   studied in Sec.~\ref{sec:wavelength}
   with (a) $\gamma=1.2$, (b) $1.05$ and (c) $1$.  The
   wavelength of initial perturbation $\lambda/\lambda_\mr{max}$ is
   $0.6$ (red), $1$ (orange), $1.5$ (green) and $2$ (blue).  }
   \label{fig:ratio_wavelength}
  \end{figure}

  \begin{figure}
    \centering
    \includegraphics[width=8cm]{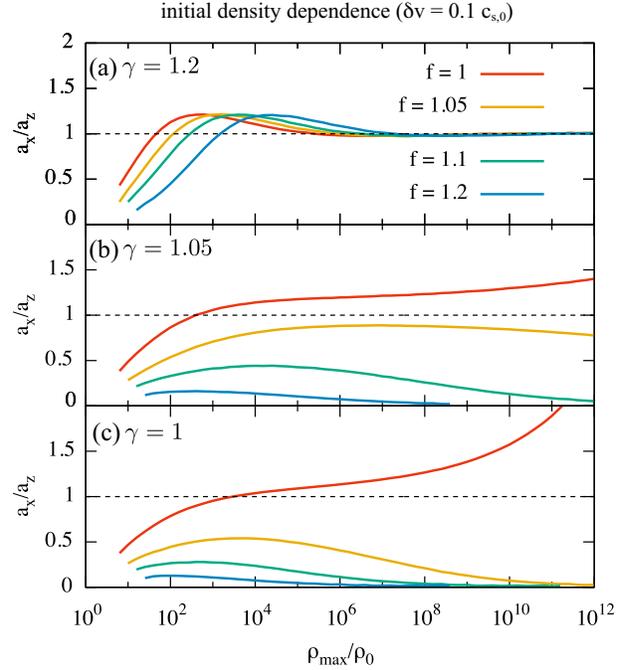}\vspace*{0.15cm} \\
    \centering
    \includegraphics[width=8cm]{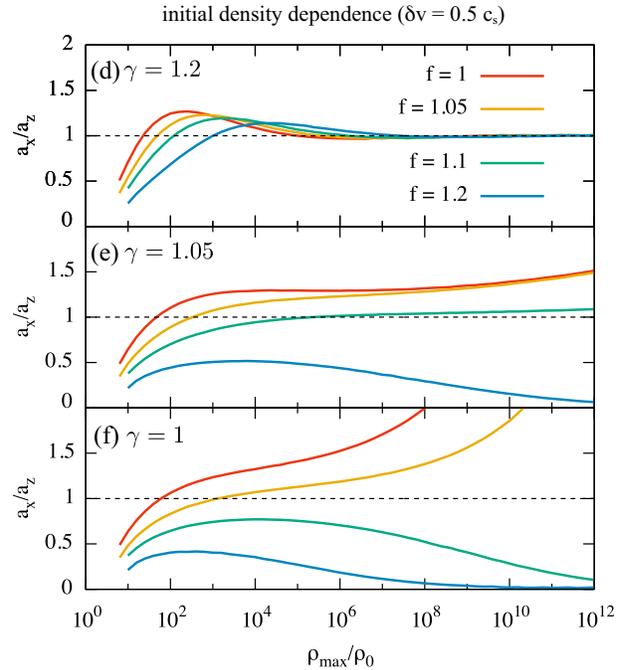}\vspace*{-0.05cm}
   \caption{Same as Fig.~\ref{fig:rat_oblt_gamma}(a) but for the models
   studied in Sec.~\ref{sec:density}
    with (a) $(\gamma,\ v_0/c_\mr{s,0})=(1.2,\ 0.1)$,
   (b) $(1.05,\ 0.1)$, (c) $(1,\ 0.1)$, (d) $(1.2,\ 0.5)$, (e) $(1.05,\
   0.5)$ and (f) $(1,\ 0.5)$.  The density enhancement factor $f$ is
   $1$ (red), $1.05$ (orange), $1.1$ (green) and $1.2$ (blue).  }
   \label{fig:ratio_density}
  \end{figure}

Having seen the cases with various amplitudes of the initial
perturbation, now we see the cases with various wavelengths of it.
Here, we present the results for the models with
$\lambda/\lambda_\mr{max}=0.6$, $1$ (fiducial), $1.5$ and $2$ (set L and
a part of set G in Table~\ref{tab:model}). We see in all models that the
filaments initially fragment into the cores. This is expected because
the linear analysis shows that modes with
$\lambda/\lambda_\mr{max}\gtrsim 0.5$ are unstable \citep[see
e.g.,][]{Larson:1985aa,Inutsuka:1992aa}.

Fig.~\ref{fig:ratio_wavelength} shows the evolution of $a_x/a_z$ for
(a) $\gamma=1.2$, (b) $1.05$ and (c) $1$. Below, we 
examine each case separately.
Firstly, for (a) $\gamma=1.2$, the $\lambda$ dependence is weak
and $a_x/a_z$ finally converges to unity irrespective of $\lambda$.
Secondly, for (b) $\gamma=1.05$, however, the $\lambda$ dependence is
strong and $a_x/a_z$ becomes larger than unity if $\lambda/\lambda_\mr{max}\le 1$ but becomes less than unity
if $\lambda/\lambda_\mr{max}\ge 1.5$ at the end of the simulations.  In
the early stage ($\rho_\mr{max}/\rho_0\lesssim 10^2$), the core collects
gases from more distant regions in the $z$ direction and thus tends to be
more prolate with larger $\lambda$.  As a result, the bar mode
begins to grow before $a_x/a_z$ reaches unity and thus the core becomes 
prolate if $\lambda/\lambda_\mr{max}\ge 1.5$,
while the bar mode begins to grow after $a_x/a_z$ exceeds unity and thus the core
becomes oblate if $\lambda/\lambda_\mr{max}\le 1$.  In the case
$\lambda/\lambda_\mr{max}= 0.6$, the core becomes spherical
because the pressure is relatively strong compared to the
gravitational force in the core with small mass.
Finally, for (c) $\gamma=1$, we see a trend similar to (b)
$\gamma=1.05$ case, although the core is more easily distorted due to the
stronger bar-mode instability.  In the case with $\lambda
= 2\,\lambda_\mr{max}$, the core is always prolate with $a_x/a_z\lesssim
0.5$ and evolves into a very elongated shape.

These results indicate that the evolution of the core strongly depends
on the wavelength of perturbation, or equivalently the interval of
fragments under our periodic boundary condition in the $z$
direction. If the wavelength is longer than that of the most unstable
mode, the core evolves differently from the fiducial case and 
tends to become prolate.

  \subsubsection{Fragmentation during radial collapse of filament}
  \label{sec:density}

Here, we present the results for filaments fragmenting during their
cylindrical radial collapse.  To induce the radial collapse of
filaments, we enhance the initial density with the density enhancement
factor $f$ (equation~\ref{eq:10}).  Meanwhile, we add a certain
amplitude of the initial velocity perturbation, to see an interplay
between the radial collapse of filaments and fragmentation \citep[see
also][]{Inutsuka:1997aa}.  If the amplitude were extremely small, the
cylindrical radial evolution would proceed too much before fragmentation
begins, i.e., the filament would collapse into the $z$ axis
($\gamma\le1$) or settle into a hydrostatic state ($\gamma>1$).  In this
section, we perform simulations with $f=1$ (fiducial), $1.05$, $1.1$ and
$1.2$ and $v_0/c_\mr{s,0}=0.1$ and $0.5$ (set D and some of set V in
Table~\ref{tab:model}).  We see in all models that filaments initially
fragment into the cores, although the fragmentation cannot be well
discriminated from the cylindrical radial collapse in the model with
$\gamma=1$, $f=1.2$ and $v_0/c_\mr{s,0}=0.1$.

Fig.~\ref{fig:ratio_density} presents the evolution of $a_x/a_z$ for
(a) $(\gamma,\ v_0/c_\mr{s,0})=(1.2,\ 0.1)$, (b) $(1.05,\
0.1)$, (c) $(1,\ 0.1)$, (d) $(1.2,\ 0.5)$, (e) $(1.05,\ 0.5)$ and (f)
$(1,\ 0.5)$, which we explain below.  
Firstly, for (a, d) $\gamma=1.2$ and $v_0/c_\mr{s,0}= 0.1$ and $0.5$, $a_x/a_z$ finally converges to unity
in all cases.
Secondly, for (b) $\gamma=1.05$ and $v_0/c_\mr{s,0}= 0.1$, however, the
overall motion of the collapsing filaments induces prolate deformation
of the core in the early phase, resulting in a more prolate shape in the
subsequent evolution with larger $f$.
Thirdly, for (d) $\gamma=1.05$ and $v_0/c_\mr{s,0}=0.5$, the trend is the
same as (b) $v_0/c_\mr{s,0}= 0.1$ but weaker, because the effect of
overall motion is less significant due to the larger flow velocity in
the $z$ direction.
Finally, for (c, f) $\gamma=1$ and $v_0/c_\mr{s,0}= 0.1$ and $0.5$, we again see a similar trend to that
seen for (b, d) $\gamma=1.05$ but with larger distortion due to the stronger
bar-mode instability, as seen in Sec.~\ref{sec:wavelength}.

In summary for this section, cylindrical radial collapse of filaments
can have a significant impact on the evolution of the cores, although
its effect is reduced if $v_0/c_\mr{s,0}$ is large.  If the filament
fragments during its radial collapse, the core tends to become prolate.

  \subsubsection{Fragmentation of filaments under external pressure}
  \label{sec:p_ext}
  \begin{figure}
   \centering \includegraphics[width=8cm]{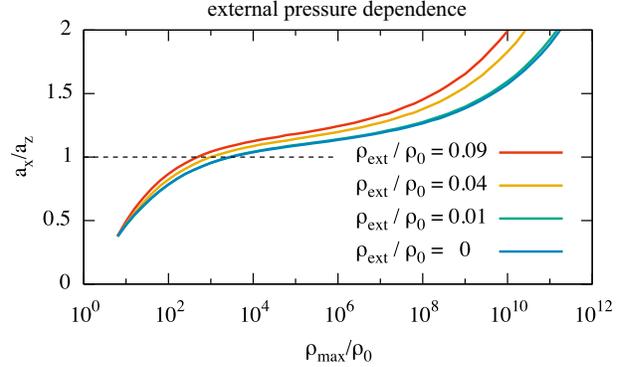} \caption{Same
   as Fig.~\ref{fig:rat_oblt_gamma}(a) but for the models studied in
   Sec.~\ref{sec:p_ext}.  The ambient gas density $\rho_\mr{ext}/\rho_0$
   is $0.09$ (red), $0.04$ (orange), $0.01$ (green) and $0$ (blue), or
   correspondingly the line mass $M_\mr{line}/M_\mr{line,cr}$ is $0.7$,
   $0.9$, $0.8$ and $1$, respectively.  .  The lines for
   $\rho_\mr{ext}/\rho_0=0.01$ and $0$ are overlapping each other.}
   \label{fig:ratio_p_ext}
  \end{figure}

Finally, we show the results for fragmentation of static filaments
under the external pressure by an ambient medium. Although we have so
far studied the ideal cases with isolated filaments, filaments are
indeed embedded in an interstellar medium with finite pressure.  Here,
we study only the cases of $\gamma=1$ and perform simulations with
$\rho_\mr{ext}/\rho_0=0$,\footnote{In practice, the density at the
boundary is taken as $\rho_\mr{b}/\rho_0=3.8\times10^{-3}$ even though
$\rho_\mr{ext}/\rho_0=0$, because of the finite size of our
computational box (see Sec.~\ref{sec:numerical}).} $0.01$, $0.04$ and
$0.09$ (set E and one of set V in Table~\ref{tab:model}).

The external pressure corresponds to the line mass of the
filaments.  The line mass is defined as
$M_\mr{line}\equiv\int_0^{r_\mr{fil}}\rho(r)2\pi r\mr{d}r$, with the
filament radius $r_\mr{fil}$ given by $\rho(r_\mr{fil})=\rho_\mr{ext}$.
The density profile of the static filament with $\gamma=1$ is given by
$\rho(r)=\rho_0(1+r^2/H_0^2)^{-2}$ and $M_\mr{line}$ takes its maximum
when the filament is isolated, i.e., $\rho_\mr{ext}=0$ and
$r_\mr{fil}=\infty$.  This maximum value, $M_\mr{line,cr}=2c_\mr{s}^2/G$
\citep[e.g.,][]{Ostriker:1964aa}, is called the critical line mass
because only sub-critical filaments can be hydrostatic and
super-critical ones are always gravitationally unstable.  The line mass
of filaments embedded in an ambient gas with finite $\rho_\mr{ext}$ is
smaller than $M_\mr{line,cr}$, as
$M_\mr{line}/M_\mr{line,cr}=1-(\rho_\mr{ext}/\rho_0)^{1/2}$ \citep[see,
e.g.,][]{Fischera:2012aa}.  Thus, the cases studied here with
$\rho_\mr{ext}/\rho_0=0$, $0.01$, $0.04$ and $0.09$ correspond to
$M_\mr{line}/M_\mr{line,cr} = 1$, $0.9$, $0.8$ and $0.7$, respectively.

 We see in all cases that the filaments fragment into the cores,
which subsequently become oblate as collapse proceeds
(Fig.~\ref{fig:ratio_p_ext}).  The dependence on the external pressure
is weak in our cases examined, although the cores tend to be more oblate
as the external pressure increases.  Such a trend can be understood as
follows.  Since $M_\mr{line}$ is smaller for larger $\rho_\mr{ext}$, as
mentioned above, the mass of fragmented cores, and hence their
gravitational potential, is also smaller.  As a result, gravitational
collapse of the cores is delayed and the cores have more time to obtain
oblate distortion in the early phase of the evolution
($\rho_\mr{max}/\rho_0\lesssim 10^3$), which explains the observed
trend.  Consistently, we also find that the time when
$\rho_\mr{max}/\rho_0$ reaches $10^{12}$ is longer for larger
$\rho_\mr{ext}$ ($t/t_\mr{ff,0}=8.3$, $8.3$, $8.5$ and $8.9$ for
$\rho_\mr{ext}/\rho_0=0$, $0.01$, $0.04$ and $0.09$, respectively). 

Here, we find that the shape of collapsing cores depends only
weakly on the external pressure.  This implies that our previous results
for isolated static filaments can be extended to the case with static
sub-critical filaments under the external pressure.  Since we have
already studied fragmentation of radially collapsing super-critical
filaments in Sec.~\ref{sec:density}, our results encompass the cases with
sub- and super-critical filaments, both of which are found in
observations \citep[see, e.g.,][]{Arzoumanian:2011aa,Fischera:2012aa}.

  \section{Conclusion and Discussion}
  \label{sec:conclusion}

We have studied the collapse of dense cores formed by fragmentation of
filaments assuming a polytropic gas with $\gamma$. By employing the
adaptive mesh refinement (AMR) technique, we are able to follow the
filament fragmentation and subsequent core collapse continuously in a
single run of simulation.  Since the self-similar spherical collapse
solution, the so-called Larson-Penston solution, is known to be unstable
due to the bar-mode instability if $\gamma<1.1$, we have focused on how
this instability affects the evolution of the cores.

We have found that the cores formed by fragmentation of filaments tend
to become spherical in the early phase of the collapse but later begin
to distort due to the bar-mode instability, if exists.  In this paper we
regard the fragmentation of an isolated hydrostatic filament triggered
by the most unstable fragmentation mode \citep[e.g.,][]{Larson:1985aa}
as a fiducial case.  For the fiducial case with $1\leq \gamma< 1.1$, we
have found the distortion becomes significant only after the central
density increases by more than ten orders of magnitude.  This is because
the distortion begins to grow out of a small seed when the background
spherical collapse becomes sufficiently close to the Larson-Penston
solution, which takes about five orders of magnitude increase in the
central density. For the fiducial case with the other $\gamma$, we see
the core becomes strongly distorted if $\gamma\leq 1$ while the core
always becomes spherical if $\gamma\geq 1.1$.  The $\gamma$ dependence
of the evolution can be understood from the fact that the bar-mode
instability exists for $\gamma<1.1$ and becomes stronger with decreasing
$\gamma$.

In addition, we have studied the filament fragmentation that occurs in a
non-fiducial way and have found the evolution of the cores can be largely
affected by the way of filament fragmentation.  The
distortion grows much faster than in the fiducial case, if the
fragmentation is triggered by a perturbation with wavelength longer
than that of the most unstable mode or proceeds during
the cylindrical radial collapse of filaments.  We caution that it is
necessary to check whether the fragmentation proceeds in the fiducial
way when applying our results for the fiducial case in an
astrophysical context.

How the filament fragmentation proceeds is determined by the initial
condition.  Theoretically, it can be addressed by simulations of
filament formation in a turbulent medium, where the perturbation is
automatically provided at the time of filament formation.  However,
although many authors have performed such simulations
\citep[e.g.,][]{Gammie:2003aa,Inoue:2012aa,
Matsumoto:2015ab,Federrath:2016aa}, none of them have focused on the
subsequent collapse of the fragmented cores.  To reduce the uncertainty
coming from the initial condition, it is important to perform a
simulation similar to this work but starting from filament formation in
future.

Let us discuss an astrophysical implication of our results on the mass
of dense cores.  As in the literature \citep[e.g.,][]{Omukai:2005aa},
the core mass at fragmentation of filaments can be estimated as follows.
In forming stars from the interstellar medium, the temperature evolution
with the increasing density draws a evolutionary path in a
density-temperature plane that is determined by environmental
conditions, such as metallicity and external radiation field \citep[see,
e.g.,][]{Omukai:2005aa,Chiaki:2016aa}.  Using the effective polytropic
index $\gamma$ defined with this path, we estimate the physical state of
the gas at each stage of the evolution. Suppose a filamentary gas
initially collapses with $\gamma<1$. It stops its radial collapse when
$\gamma$ exceeds unity, because the critical $\gamma$ for filaments is
$\gamma_\mr{cr}=1$, i.e., the pressure increases more (less) rapidly than
the gravitational force if $\gamma>1$ ($\gamma<1$). Then, the
pressure-supported filament fragments into dense cores
\citep{Tsuribe:2006aa}. The cores subsequently collapse and form stars
inside.  Here, the mass of the fragments can be estimated as the Jeans
mass for the density and temperature when $\gamma$ exceeds unity
\citep[e.g.,][]{Larson:1985aa,Inutsuka:1992aa} and gives a good estimate
for the initial mass of the cores.

The final mass of the star-forming cores, however, can be largely
altered if the distortion of collapsing cores results in their
re-fragmentation.  We have shown for the fiducial case that it takes
more than ten orders-of-magnitude in the density increase for the
distortion to become non-linear.  Thus, such re-fragmentation is not
likely in most cases, because the phase with $\gamma<1.1$ does not last
such long.  This can happen, however, in the following situations in the
early Universe: in supermassive ($\sim 10^5\,M_\odot$) star formation in
strongly irradiated pristine clouds
\citep[e.g.,][]{Omukai:2001aa,Bromm:2003aa,Sugimura:2014aa}, the gas
collapses with almost constant temperature of $\sim10^4\cmr{K}$ due to
the Ly$\alpha$ cooling; in Pop II star formation in the very
high-redshift ($z\gtrsim 20$) Universe, the gas evolves with the
temperature of the cosmic microwave background at that time
\citep[e.g.,][]{Omukai:2005aa,Safranek-Shrader:2014ab}.  In these
exceptional cases, the cores can be significantly distorted and finally
re-fragment before forming stars. In addition, it should be emphasized
again that in non-fiducial cases, i.e., if the filament is not
hydrostatic or fragmentation is not triggered by the most-unstable mode,
the re-fragmentation of the cores can be important.

Here, we give qualitative estimate of the condition for
re-fragmentation, although its numerical investigation is out of the
scope of this work.  Suppose that the collapse of distorted cores is
delayed at some moment, possibly due to the increase of $\gamma$.  In
such a case, a rough estimate can be made using the critical wavelengths
for unstable modes of static filaments and sheets, about four and six
times the scale length, respectively \citep[e.g.,][]{Larson:1985aa}.  If
the axial ratio of the cores is larger than twice the ratio of the
critical wavelengths to the scale length, i.e., eight for prolate cores
and twelve for oblate ones, they are able to fragment into more than two
depending on the initial amplitude of the unstable modes.  For more
realistic estimate, however, numerical simulations dedicated to the
re-fragmentation of distorted cores are needed.

The initial condition dependence found in this work suggests an
observational relation between the physical state of filaments and the
shapes of the cores within them
\citep[e.g.,][]{Myers:1991aa,Ryden:1996aa}.  We suggest that the cores
tend to be more prolate along the filaments if the interval of the cores is
longer than the wavelength of the most-unstable mode, i.e., four times
the diameter of the filaments \citep[e.g.,][]{Inutsuka:1992aa}, or the
filaments show the sign of overall cylindrical radial collapse.  These
relations can be observationally tested.

In the current calculation, we have neglected the effects of a magnetic
field and a rotational and turbulent velocity field, in order to extract
only the effect of the bar-mode instability on the evolution of cores.
We should ultimately account for them in studying the evolution of the
dense cores formed from filaments, as it has been suggested that a
rotational velocity field \citep[e.g.,][]{Matsumoto:1997aa} and a
magnetic field \citep[e.g.,][]{Nakamura:1993aa} affect the evolution of
cores. It is also known that the turbulence generated during
gravitational collapse of cores can play an important role in the
evolution of the cores \citep[e.g.,][]{Federrath:2011aa}.  We would like
to address these issues in future publication.

\section*{Acknowledgements}

The authors would like to thank Gen Chiaki, Shu-ichiro Inutsuka,
Kazunari Iwasaki, Sanemichi Takahashi and Toru Tsuribe for fruitful
discussions.  The numerical simulations were performed on the Cray XC30
at CfCA of the National Astronomical Observatory of Japan.  This work is
supported in part by MEXT/JSPS KAKENHI Grant Number 15J03873 (KS),
26400233, 26287030 and 24244017 (TM) and 25287040 (KO).


\begin{thebibliography}{}
\expandafter\ifx\csname natexlab\endcsname\relax\def\natexlab#1{#1}\fi

\bibitem[{{Andr{\'e}} {et~al.}(2010){Andr{\'e}}, {Men'shchikov}, {Bontemps},
  {K{\"o}nyves}, {Motte}, {Schneider}, {Didelon}, {Minier}, {Saraceno},
  {Ward-Thompson}, {di Francesco}, {White}, {Molinari}, {Testi}, {Abergel},
  {Griffin}, {Henning}, {Royer}, {Mer{\'{\i}}n}, {Vavrek}, {Attard},
  {Arzoumanian}, {Wilson}, {Ade}, {Aussel}, {Baluteau}, {Benedettini},
  {Bernard}, {Blommaert}, {Cambr{\'e}sy}, {Cox}, {di Giorgio}, {Hargrave},
  {Hennemann}, {Huang}, {Kirk}, {Krause}, {Launhardt}, {Leeks}, {Le Pennec},
  {Li}, {Martin}, {Maury}, {Olofsson}, {Omont}, {Peretto}, {Pezzuto}, {Prusti},
  {Roussel}, {Russeil}, {Sauvage}, {Sibthorpe}, {Sicilia-Aguilar}, {Spinoglio},
  {Waelkens}, {Woodcraft}, \& {Zavagno}}]{Andre:2010aa}
{Andr{\'e}}, P., {Men'shchikov}, A., {Bontemps}, S., {et~al.} 2010, \aap, 518,
  L102

\bibitem[{{Arzoumanian} {et~al.}(2011){Arzoumanian}, {Andr{\'e}}, {Didelon},
  {K{\"o}nyves}, {Schneider}, {Men'shchikov}, {Sousbie}, {Zavagno}, {Bontemps},
  {di Francesco}, {Griffin}, {Hennemann}, {Hill}, {Kirk}, {Martin}, {Minier},
  {Molinari}, {Motte}, {Peretto}, {Pezzuto}, {Spinoglio}, {Ward-Thompson},
  {White}, \& {Wilson}}]{Arzoumanian:2011aa}
{Arzoumanian}, D., {Andr{\'e}}, P., {Didelon}, P., {et~al.} 2011, \aap, 529, L6

\bibitem[{{Bromm} \& {Loeb}(2003)}]{Bromm:2003aa}
{Bromm}, V., \& {Loeb}, A. 2003, \apj, 596, 34

\bibitem[{{Chiaki} {et~al.}(2016){Chiaki}, {Yoshida}, \&
  {Hirano}}]{Chiaki:2016aa}
{Chiaki}, G., {Yoshida}, N., \& {Hirano}, S. 2016, \mnras, 463, 2781

\bibitem[{{Clarke} {et~al.}(2016){Clarke}, {Whitworth}, \&
  {Hubber}}]{Clarke:2016ac}
{Clarke}, S.~D., {Whitworth}, A.~P., \& {Hubber}, D.~A. 2016, \mnras, 458, 319

\bibitem[{{Federrath}(2016)}]{Federrath:2016aa}
{Federrath}, C. 2016, \mnras, 457, 375

\bibitem[{{Federrath} {et~al.}(2011){Federrath}, {Sur}, {Schleicher},
  {Banerjee}, \& {Klessen}}]{Federrath:2011aa}
{Federrath}, C., {Sur}, S., {Schleicher}, D.~R.~G., {Banerjee}, R., \&
  {Klessen}, R.~S. 2011, \apj, 731, 62

\bibitem[{{Fischera} \& {Martin}(2012)}]{Fischera:2012aa}
{Fischera}, J., \& {Martin}, P.~G. 2012, \aap, 542, A77

\bibitem[{{Gammie} {et~al.}(2003){Gammie}, {Lin}, {Stone}, \&
  {Ostriker}}]{Gammie:2003aa}
{Gammie}, C.~F., {Lin}, Y.-T., {Stone}, J.~M., \& {Ostriker}, E.~C. 2003, \apj,
  592, 203

\bibitem[{{Gritschneder} {et~al.}(2017){Gritschneder}, {Heigl}, \&
  {Burkert}}]{Gritschneder:2017aa}
{Gritschneder}, M., {Heigl}, S., \& {Burkert}, A. 2017, \apj, 834, 202

\bibitem[{{Hanawa} \& {Matsumoto}(2000)}]{Hanawa:2000aa}
{Hanawa}, T., \& {Matsumoto}, T. 2000, \pasj, 52, 241

\bibitem[{{Heigl} {et~al.}(2016){Heigl}, {Burkert}, \& {Hacar}}]{Heigl:2016aa}
{Heigl}, S., {Burkert}, A., \& {Hacar}, A. 2016, \mnras, 463, 4301

\bibitem[{{Heitsch}(2013{\natexlab{a}})}]{Heitsch:2013aa}
{Heitsch}, F. 2013{\natexlab{a}}, \apj, 769, 115

\bibitem[{{Heitsch}(2013{\natexlab{b}})}]{Heitsch:2013ab}
---. 2013{\natexlab{b}}, \apj, 776, 62

\bibitem[{{Inoue} \& {Inutsuka}(2012)}]{Inoue:2012aa}
{Inoue}, T., \& {Inutsuka}, S.-i. 2012, \apj, 759, 35

\bibitem[{{Inutsuka} \& {Miyama}(1992)}]{Inutsuka:1992aa}
{Inutsuka}, S.-I., \& {Miyama}, S.~M. 1992, \apj, 388, 392

\bibitem[{{Inutsuka} \& {Miyama}(1997)}]{Inutsuka:1997aa}
{Inutsuka}, S.-i., \& {Miyama}, S.~M. 1997, \apj, 480, 681

\bibitem[{{Lai}(2000)}]{Lai:2000aa}
{Lai}, D. 2000, \apj, 540, 946

\bibitem[{{Larson}(1969)}]{Larson:1969aa}
{Larson}, R.~B. 1969, \mnras, 145, 271

\bibitem[{{Larson}(1985)}]{Larson:1985aa}
---. 1985, \mnras, 214, 379

\bibitem[{{Machida} \& {Matsumoto}(2012)}]{Machida:2012aa}
{Machida}, M.~N., \& {Matsumoto}, T. 2012, \mnras, 421, 588

\bibitem[{{Matsumoto}(2007)}]{Matsumoto:2007aa}
{Matsumoto}, T. 2007, \pasj, 59, 905

\bibitem[{{Matsumoto} {et~al.}(2015){Matsumoto}, {Dobashi}, \&
  {Shimoikura}}]{Matsumoto:2015ab}
{Matsumoto}, T., {Dobashi}, K., \& {Shimoikura}, T. 2015, \apj, 801, 77

\bibitem[{{Matsumoto} \& {Hanawa}(1999)}]{Matsumoto:1999aa}
{Matsumoto}, T., \& {Hanawa}, T. 1999, \apj, 521, 659

\bibitem[{{Matsumoto} {et~al.}(1997){Matsumoto}, {Hanawa}, \&
  {Nakamura}}]{Matsumoto:1997aa}
{Matsumoto}, T., {Hanawa}, T., \& {Nakamura}, F. 1997, \apj, 478, 569

\bibitem[{{Myers} {et~al.}(1991){Myers}, {Fuller}, {Goodman}, \&
  {Benson}}]{Myers:1991aa}
{Myers}, P.~C., {Fuller}, G.~A., {Goodman}, A.~A., \& {Benson}, P.~J. 1991,
  \apj, 376, 561

\bibitem[{{Nagasawa}(1987)}]{Nagasawa:1987aa}
{Nagasawa}, M. 1987, Progress of Theoretical Physics, 77, 635

\bibitem[{{Nakamura}(2000)}]{Nakamura:2000aa}
{Nakamura}, F. 2000, \apj, 543, 291

\bibitem[{{Nakamura} {et~al.}(1993){Nakamura}, {Hanawa}, \&
  {Nakano}}]{Nakamura:1993aa}
{Nakamura}, F., {Hanawa}, T., \& {Nakano}, T. 1993, \pasj, 45, 551

\bibitem[{{Omukai}(2001)}]{Omukai:2001aa}
{Omukai}, K. 2001, \apj, 546, 635

\bibitem[{{Omukai} {et~al.}(2005){Omukai}, {Tsuribe}, {Schneider}, \&
  {Ferrara}}]{Omukai:2005aa}
{Omukai}, K., {Tsuribe}, T., {Schneider}, R., \& {Ferrara}, A. 2005, \apj, 626,
  627

\bibitem[{{Ostriker}(1964)}]{Ostriker:1964aa}
{Ostriker}, J. 1964, \apj, 140, 1056

\bibitem[{{Penston}(1969)}]{Penston:1969ab}
{Penston}, M.~V. 1969, \mnras, 145, 457

\bibitem[{{Roy} {et~al.}(2015){Roy}, {Andr{\'e}}, {Arzoumanian}, {Peretto},
  {Palmeirim}, {K{\"o}nyves}, {Schneider}, {Benedettini}, {Di Francesco},
  {Elia}, {Hill}, {Ladjelate}, {Louvet}, {Motte}, {Pezzuto}, {Schisano},
  {Shimajiri}, {Spinoglio}, {Ward-Thompson}, \& {White}}]{Roy:2015aa}
{Roy}, A., {Andr{\'e}}, P., {Arzoumanian}, D., {et~al.} 2015, \aap, 584, A111

\bibitem[{{Ryden}(1996)}]{Ryden:1996aa}
{Ryden}, B.~S. 1996, \apj, 471, 822

\bibitem[{{Safranek-Shrader} {et~al.}(2014){Safranek-Shrader},
  {Milosavljevi{\'c}}, \& {Bromm}}]{Safranek-Shrader:2014ab}
{Safranek-Shrader}, C., {Milosavljevi{\'c}}, M., \& {Bromm}, V. 2014, \mnras,
  440, L76

\bibitem[{{Stod{\'o}lkiewicz}(1963)}]{Stodolkiewicz:1963aa}
{Stod{\'o}lkiewicz}, J.~S. 1963, \actaa, 13, 30

\bibitem[{{Sugimura} {et~al.}(2014){Sugimura}, {Omukai}, \&
  {Inoue}}]{Sugimura:2014aa}
{Sugimura}, K., {Omukai}, K., \& {Inoue}, A.~K. 2014, \mnras, 445, 544

\bibitem[{{Truelove} {et~al.}(1997){Truelove}, {Klein}, {McKee}, {Holliman},
  {Howell}, \& {Greenough}}]{Truelove:1997aa}
{Truelove}, J.~K., {Klein}, R.~I., {McKee}, C.~F., {et~al.} 1997, \apjl, 489,
  L179

\bibitem[{{Tsuribe} \& {Inutsuka}(1999)}]{Tsuribe:1999aa}
{Tsuribe}, T., \& {Inutsuka}, S.-i. 1999, \apj, 526, 307

\bibitem[{{Tsuribe} \& {Omukai}(2006)}]{Tsuribe:2006aa}
{Tsuribe}, T., \& {Omukai}, K. 2006, \apjl, 642, L61

\bibitem[{{Yahil}(1983)}]{Yahil:1983aa}
{Yahil}, A. 1983, \apj, 265, 1047

\end{thebibliography}


  \appendix

  \section{Resolution check}
  \label{sec:resolution_dependence}

To check the resolution dependence of our results, here we repeat the
simulation shown in Sec.~\ref{sec:time_evolution} ($\gamma=1.05$ model
of set G in Table~\ref{tab:model}) but with different resolutions.  In
our AMR simulation, there are two parameters controlling the resolution:
the initial number of meshes in each direction, $N_\mr{ini}$, and the
minimum number of meshes per one Jeans length $\lambda_\mr{J}$
(equation~\ref{eq:3}), $N_\mr{ref}$.  The former controls the resolution
during the initial fragmentation phase while the latter does during the
subsequent collapse phase.  We take $N_\mr{ini}=256$ and $N_\mr{ref}=32$
as the fiducial resolution in this paper, but here we change
$N_\mr{ini}$ to $128$, $512$ or $1024$ or $N_\mr{ref}$ to $8$, $16$ or
$64$.

\begin{figure}
   \centering \includegraphics[width=8cm]{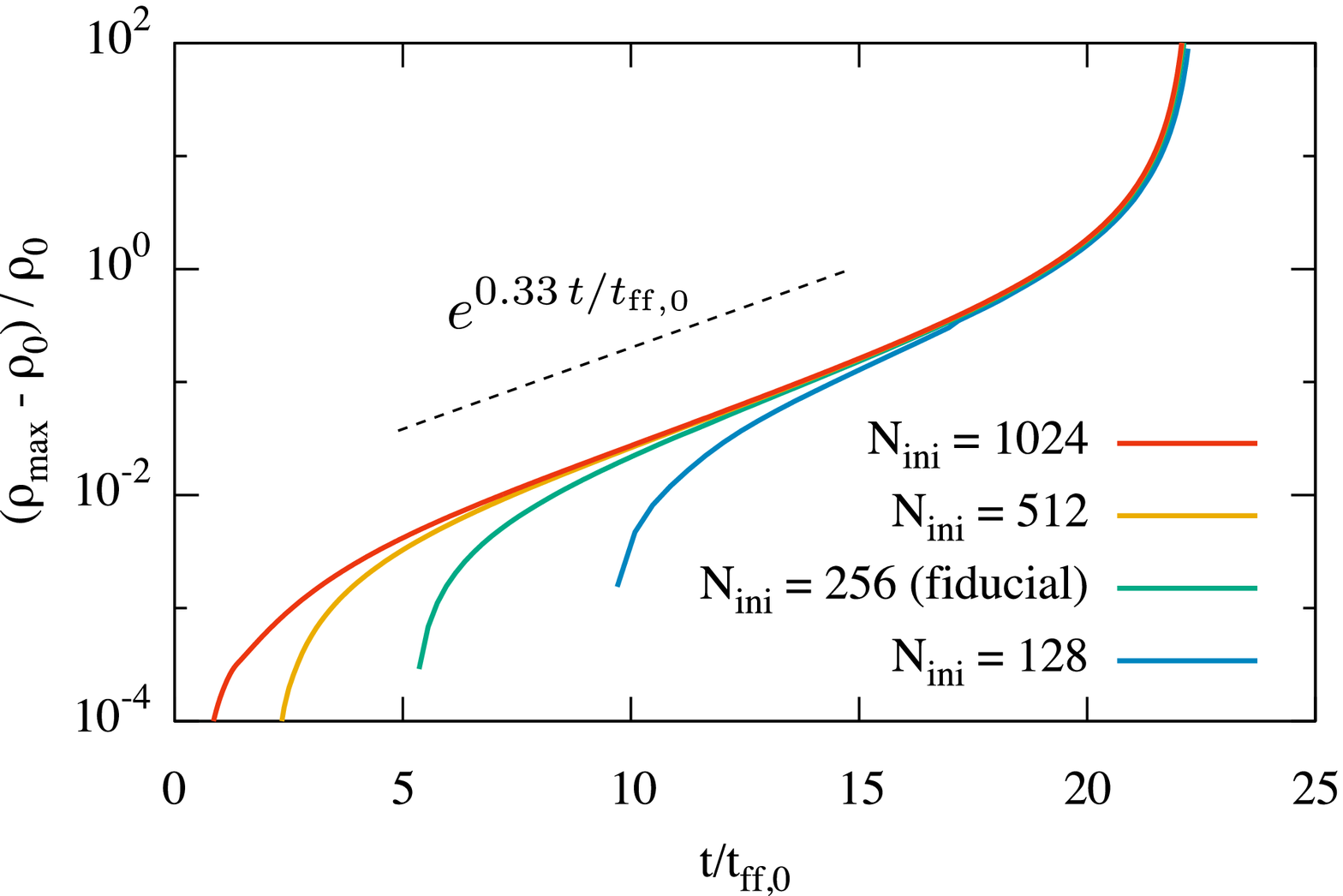}
   \caption{Same as Fig.~\ref{fig:rhomax_g105} but with $N_\mr{ini}$
   taken as $1024$ (red), $512$ (orange), $256$ (green; fiducial) and
   $128$ (blue).  } \label{fig:rhomax_resolution}
  \end{figure}

Fig.~\ref{fig:rhomax_resolution} shows the $N_\mr{ini}$ dependence of
the time evolution of $\delta=(\rho_\mr{max}-\rho_0)/\rho_0$
(equation~\ref{eq:5}). We see that the evolution of $\delta$ can be
correctly followed from the earlier stage by adopting larger
$N_\mr{ini}$, although $N_\mr{ini}$ hardly affects the evolution at the
time of fragmentation.  With fiducial $N_\mr{ini}=256$, the evolution
is reliable after $\delta \gtrsim 10^{-2}$.

\begin{figure}
   \centering
   \includegraphics[width=8cm]{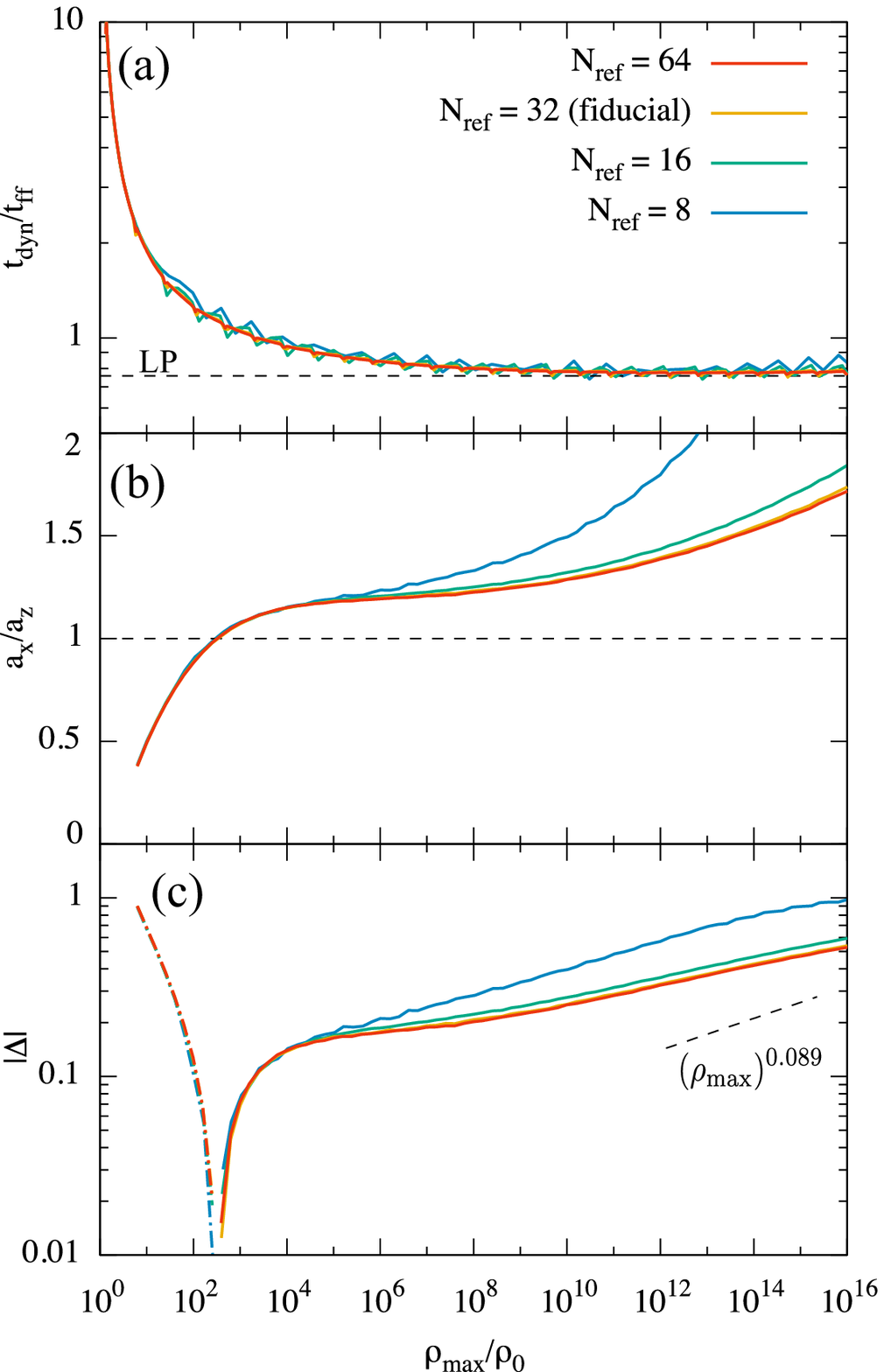}
   \caption{Same as Fig.~\ref{fig:tdynff_rat_oblt_g105} but with
   $N_\mr{ref}$ taken as $64$ (red), $32$ (orange; fiducial), $16$
   (green; fiducial) and $8$ (blue). The lines for $N_\mr{ref}=64$ and
   $32$ are overlapped and difficult to be separately seen.}
   \label{fig:tdynff_rat_oblt_resolution}
  \end{figure}

We plot the $N_\mr{ref}$ dependence of (a) $t_\mr{dyn}/t_\mr{ff}$ (see
above equation~\ref{eq:9}), (b) $a_x/a_z$ (equation~\ref{eq:6}) and (c)
$\Delta=2(a_x-a_z)/(a_x+a_z)$ (equation~\ref{eq:19}) in
Fig.~\ref{fig:tdynff_rat_oblt_resolution}.  We see in panel (a) that the
features of $t_\mr{dyn}/t_\mr{ff}$ appearing every four-time density
increase are generated by numerical errors at refinement, which can be
suppressed by adopting larger $N_\mr{ref}$.  These features, however,
affect the overall evolution only slightly.  Panels (b) and (c) show the
$N_\mr{ref}$ dependence of the growth of the bar-mode instability. We
see that the $N_\mr{ref}$ dependence is substantial for $N_\mr{ref}=8$,
but becomes weaker for $N_\mr{ref}=16$ and cannot be seen by eyes for
$N_\mr{ref}\ge 32$.  This suggests that the minimum resolution is
$N_\mr{ref}\sim 16$ and that our fiducial resolution with
$N_\mr{ref}=32$ is sufficient.

Let us discuss the minimum resolution required to correctly solve the
dynamics of self-gravitating gas.  We here obtain the condition
$N_\mr{ref}\ge 16$, required to correctly follow the growth of the bar
mode.  The most often-used condition in the literature is the so-called
Truelove condition, $N_\mr{ref}\ge 4$ \citep{Truelove:1997aa}, which is
required to avoid artificial fragmentation.  More recently,
\cite{Federrath:2011aa} suggest to use a condition $N_\mr{ref}\ge 32$ to
resolve the turbulence generated during gravitational collapse of cores.
Note that these three condition are derived to follow the different
physical processes.  In performing simulations, either one of the above
three conditions should be used depending on the process to be followed.

\end{document}